\documentclass[preprint,12pt]{elsarticle}

\usepackage{hyperref}
\usepackage{longtable}
\usepackage{graphics}
\usepackage{graphicx}
\usepackage{epsfig}
\usepackage{amssymb}
\usepackage{amsthm}
\usepackage{lineno}
\usepackage{setspace}
\usepackage{lscape}
\usepackage{xcolor}
\usepackage{lscape}
\usepackage{graphics}
\usepackage{graphicx}
\usepackage{epsfig}
\usepackage{amssymb}
\usepackage{amsthm}
\usepackage{setspace}
\usepackage{lscape}
\usepackage{xcolor}
\usepackage{amsmath}
\usepackage{makecell}

\journal{}

\begin{document}

\begin{frontmatter}

\title{A time-fractional dual-phase-lag framework to investigate transistors with TMTC channels (TiS$_3$, In$_4$Se$_3$) and size-dependent properties}

\author[2]{Mohammad Hosein Fotovvat}
\author[1]{Zahra Shomali\corref{cor1}}
\cortext[cor1]{Corresponding author, Tel.: +982182884785}
\ead{shomali@modares.ac.ir}
\address[2]{Mechanical Engineering Department, Faculty of Engineering, University of Zanjan, Zanjan, Iran}
\address[1]{Department of Physics, Faculty of Basic Sciences, Tarbiat Modares University, Tehran, Iran}

\begin{abstract}
In this study, a time fractional dual-phase-lag model with temperature jump boundary condition as a choice for the Fourier's law replacement in thermal modeling of transistors, is utilized. In more details, the numerical simulation of heat transfer in newly proposed TMTC field effect transistors using fractional DPL equation has been investigated. Moreover, the Caputo fractional derivative is employed to formulate the finite difference scheme for discretization of the fractional DPL model. In order to obtain more precise results for the peak temperature rise, the temperature and heat flux profiles, the size-dependent thermal properties are taken into account. Also, the temperature jump boundary condition has been also applied by means of a mixed-type boundary condition. It is obtained that considering size-dependent thermal characteristics for transistors under study, results in an increase of the peak temperature rise. As an instance, considering the film dependent thermal properties for a 1-D silicone MOSFET with uniform heat generation at the upper part, makes the maximum temperature increase up to 250 percent. Furthermore, considering constant bulk thermal properties for the silicon MOSFET, certain oscillations are observed in the time-variation of the peak temperature rise for $\alpha$= 0.7, 0.9 and 1. This presents the so-called negative bias temperature instability appearing in electronic nano-semiconductor devices. Finally, the hotspot temperature has been researched in transistors containing two-dimensional materials with quasi one-dimensional band structure channels. It is obtained that among the studied FETs, titanium trisulfide with maximum temperature increase of 19.63 K exhibits the least peak temperature rise. This presents that TiS$_3$ may be an acceptable silicon channel replacement as far as the thermal issues are concerned.

\end{abstract}


\end{frontmatter}
\section{Introduction}
The necessity for higher speed in constant power density, has decreased the size of the MOSFETs. New generation of transistors are built in sizes of less than 10 nm, in order to have more efficiency than that of the silicon-based transistors \cite{galiybuilding,Superlattice2018}. On the contrary, as the size of the MOSFETs shrinks, they encounter imperfections and problems which disrupt their performance. Two-dimensional materials like graphene and transition metal dichalcogenide (TMD) with chemical composition of MX$_2$ (for example MoS$2$), due to their reduced leakage current, are suitable candidates for use as the channel in field effect transistors \cite{Shomali2019A}. However, excessive reduction of the channel dimensions in these transistors makes materials such as graphene to lose their perfect properties such as low weight and high mobility. Put differently, for two dimensional materials like graphene and transition metal dichalcogenides, reducing the system size to 10 nm or less, increases the edge effects and phonon edge scattering notably. These effects ruin the performance of the transistors \cite{wimmer2008spin}. With an eye toward solving this problem, one can utilize quasi one-dimensional materials for which the phonon scattering effect at edges is negligible. The two dimensional transition metal trichalcogenide materials (TMTCs) \cite{TMTC} with chemical composition of MX$_3$ such as TiS$3$ and In$_4$X$_3$ as an instance In$_4$Se$_3$, have quasi one-dimensional band structures and their energy gaps are about 5 ev. These quasi-one-dimensional materials have many optoelectronic properties of two-dimensional substances and also in addition has an advantage of imperceptible edge scattering effect \cite{lipatov2018quasi}. Consequently, building a transistor with size less than 10 nm and higher reliability are more probable when one uses TMTCs instead of TMDCs or graphene-like materials.

On the other hand, as the length scale of the electronic devices shrinks to the nanoscale, the classical equations of continuum mechanics fail to endorse the experimental data. Hence, it is necessary to develop creative models which predict more accurate results relative to the classical equations such as Fourier's law with less computational cost and more simplicity \cite{Nasri2015,Nasri20152}. Meanwhile, the dual-phase-lag (DPL) model \cite{tzou1995unified,tzou2001temperature} as a constitutive equation to investigate the non-Fourier heat transport in different cases of nanoscale MOSFETs has drawn attentions \cite{ghazanfarianshomali2012investigation,ghazanfarianabbassi2012investigation,samian2013thermal,samian2014transient,Moghaddam2014,shomali2014investigation,shomali2015effect,Rev2015,saghatchi2015,Shomali2016,Shiomi2016,Nasri2017,Pedar2017,shomali2017cht,ghazanfarian2018,Zobiri2021}. As a consequence of the easiness, straightforwardness, adaptability, and preciseness of this method, it is a reliable choice for heat transfer investigation of transistors with disparate geometries. On the other hand, in recent years, the subject concerning combination of the fractional calculus and different diffusive models in order to investigate the anomalous diffusive phenomena considering the non-local, has been the center of attention. Indeed, fractional calculus has been successfully utilized for the modification of many existing models of physical phenomena. Actually, the theory of fractional derivatives and integral has been developed in the nineteenth century. Caputo and Mainardi have made use of fractional derivatives and found an adequate agreement between the obtained results and what has been found from the experiment \cite{Caputo1971,Caputo1971-2,Caputo1974}. In 2010, Sherief \emph{et al.} presented the fractional single-phase-lagging for the sake of viscoelastic material behavior using the fractional calculus and Caputo fractional derivative. A satisfactory consistency between the numerical results obtained from the new suggested model and the available experimental data is seen \cite{sherief2010fractional}. Also, Mishra and Rai have formulated the fractional single-phase-lagging (FSPL) heat conduction equation via employing the single-phase lag model and Caputo fractional derivative \cite{mishra2016numerical}. In this study, the discretization has been performed using the compact difference scheme. Further, the non-Fourier heat conduction within a finite thin film under the effect of a time-varying and spatially-decaying laser heating considering the Dirichlet Boundary condition, has been investigated. In 2018, Ji \emph{et al.} have established a fractional dual phase lag model based on the finite difference numerical method, to investigate the heat conduction in nanoscale devices \cite{ji2018numerical2}. Also, they have investigated the non-Fourier heat transport in a thin two-layer film exposed to the ultrashort-pulsed laser heating applying the fractional DPL model \cite{ji2019numerical}.

Here, in order to investigate reliability and functionality of the titanium trisulfide (TiS$_3$) and tetraindium triselenide (In$_4$Se$_3$) quasi one-dimensional transistors, as an alternative to the previous generation silicon transistors, temperature distribution profile and the maximum temperature increase in these transistors are investigated using the fractional DPL model. The heat transport through a silicon slab considering the temperature jump boundary condition, and in existence of the heat generation zone, is scrutinized. In the present study, the heat conduction in MOSFETs will be investigated. In more details, first, the one-dimensional silicon MOSFETs will be investigated and the results will be used for verification. Then, the MOSFET containing silicon channel and buried oxide will be modeled using the fractional DPL. Then, the peak temperature rise and the temperature profile for quasi one-dimensional MOSFET transistors including newly proposed transition metal trichalcogenides gates, TiS$_3$ and In$_4$Se$_3$, are studied using the fractional dual phase lag model. The MOSFET with maximum temperature increase of 19.63 K is found to have the least temperature rise. So, it can be suggested as the suitable replacement of the one-dimensional silicon channel. Moreover, the effect of size-dependent thermal conductivity on thermal behavior of the quasi-one-dimensional TMTC MOSFETs is studied.

The structure of present study is as follows. In Sec. \ref{Mathematical modeling}, the governing equations, and the assumptions will be demonstrated. The numerical method is exhaustively discussed in Sec. \ref{Solution technique}. The validation results for different cases and the obtained results are given in Secs. \ref{verification} and \ref{results}. At last, the paper is concluded in Sec. \ref{con}.

\section{Mathematical modeling}
\label{Mathematical modeling}
In this section, first, the derivation of fractional dual phase lag equation is presented. Then, the discretization of the governing equation, the way of applying the boundary condition, the initial conditions and the numerical finite difference scheme for discretization of the fractional DPL equation will be presented.

\subsection{fractional DPL}
The DPL model for the one-dimensional geometry is written as,
\begin{equation} \label{DPL_Eq_Scalar}
	q(x,t+\tau_q) = -kT_x(x,t+\tau_t).
\end{equation}

Where the $q$ is the heat flux and T$_x$ is the temperature gradient vector. The thermal conductivity is also presented as $k$. On the other hand, the differential relation for energy conservation law is:
\begin{equation} \label{Energy_eq}
	\rho C_pT_t(x,t)=-q_x(x,t)+S(x,t).
\end{equation}

Here $\rho$, $C_p$ and $S(x,t)$, are respectively, the mass density, the specific heat and the heat source. There are points on how the heat source forms during the self-heating in a MOSFET. This electric field makes the charge carriers accelerate. So, they may scatter from each other, phonons, imperfections and interfaces. The electrons discharge energy, scattering from the phonons, while self-heating the MOSFET throughout the prominent Joule heating mechanism. Accordingly, the Joule heating results in a local temperature rise, hot spot. The local temperatures are notably higher than the average temperature. Joule heating is evidently the dominant self-heating process for low-dimensional systems, among all other heat generation mechanisms including current crowding and thermo-electrics.

In nano-electronics, the reliability is defined by the hottest region temperature of the die. So, hotspots formation mainly manages the higher-level packaging and thermal management solutions like the material selections and the design of the heat spreaders. Consequently, the nano-device and the die thermal demeanor has an important role in detecting cooling requirements and also finding the environmental effects. It means, when the temperature gets larger due to the power density increase, the performance of the nano-device will be restricted. Therefore, many works have been done to make the cooling techniques optimized for heat spreading enhancement. One can say that the channels built from materials which produce the minimum peak temperature rise are the most suitable and desired choices for the nano-electronics industry. Rightfully so, maintaining the temperature of the nano-device under the stated temperature threshold is easier for the ones with lower maximum temperatures. Subsequently, the operation of the transistors in digital devices such as laptops, tablets, and cellphones completely relies on their thermal behavior. In more detail, inefficient thermal operation of these transistors causes the temperature to increase too much. This augmentation makes the nano-device stop working perpetually. The nanoscale heat transfer can be investigated via DPL model using three methods. In the first method, the DPL equation is expanded using the Taylor series and then the heat flux is replaced utilizing the relation dragged out from the energy equation. Consequently, an equation containing only temperature as the unknown parameter is obtained \cite{ghazanfarian2009effect,ghazanfarianshomali2012investigation,dai2013accurate}. Also, the unexpanded DPL and the energy equation can be simultaneously solved \cite{dai2002approximate}. Dealing with the third method, the expanded DPL equation and the energy equation are together worked out to calculate the temperature profile and the heat flux \cite{shomali2015effect}. When the thermal properties are temperature dependent, the equations are non-linear and one should use the third method. Here, the first method which considers the fractional Taylor expansion of the heat flux and the temperature gradient will be used:

\begin{align}
	\nonumber
	&q(x,t+\tau_q) = q(x,t)+\frac{(\tau_q)^\alpha}{\Gamma(1+\alpha)}{}_{0}^{C}\textrm{\textit{D}}_t^\alpha q(x,t)+\cdots\\
	\label{T_TSE}
	&T_x(x,t+\tau_t) = T_x(x,t)+\frac{(\tau_t)^\alpha}{\Gamma(1+\alpha)}{}_{0}^{C}\textrm{\textit{D}}_t^\alpha T_x(x,t)+\cdots.
\end{align}
In the above equations, $0<\alpha<1$ and the $^{C}D^{\alpha}_{t}$ operator is the Caputo fractional derivative of order $\alpha$ \cite{caputo1967linear}:
\begin{equation}
	\label{Caputo_Fractional_Derivative}
	_{0}^{C}\textrm{\textit{D}}_t^\alpha f(t) = \frac{1}{\Gamma(1-\alpha)}\int_{0}^{t}\frac{{f}'(\xi)}{(t-\xi)^\alpha}d\xi.
\end{equation}
By substituting Eqs. (\ref{T_TSE}) in Eq. (\ref{DPL_Eq_Scalar}), one can find the following relation:
	\begin{align}
		\label{qT_TSE}
		q(x,t)+\frac{(\tau_q)^\alpha}{\Gamma(1+\alpha)}{}_{0}^{C}\textrm{\textit{D}}_t^\alpha q(x,t)=
		-k\left(T+\frac{(\tau_t)^\alpha}{\Gamma(1+\alpha)}{}_{0}^{C}\textrm{\textit{D}}_t^\alpha T\right)_x(x,t).
	\end{align}
	Taking the x-derivative of two sides of Eq. (\ref{DPL_Eq_Scalar}) along using the energy conversation law, the fractional DPL is obtained as:	
	\begin{align} \label{FDPL_eq}
		\rho C_p\left[T_t+\frac{(\tau_q)^\alpha}{\Gamma(1+\alpha)}{}_{0}^{C}\textrm{\textit{D}}_t^\alpha T_t\right](x,t)&=k\left(T+\frac{(\tau_t)^\alpha}{\Gamma(1+\alpha)}{}_{0}^{C}\textrm{\textit{D}}_t^\alpha T\right)_{xx}(x,t) \nonumber\\
		&+f(x,t).
	\end{align}
	Here, $f(x,t)$ is defined as
		\begin{equation} \label{Source_Term}
		f(x,t)=S(x,t)+\frac{(\tau_q)^\alpha}{\Gamma(1+\alpha)}{}_{0}^{C}\textrm{\textit{D}}_t^\alpha S(x,t).
	\end{equation}

\subsection{Non-dimensional governing equations}
	In order to make the governing equations (\ref{FDPL_eq}) and (\ref{Source_Term}) non-dimensional, the following non-dimensional parameters are defined:

	\begin{align} \label{Dimensionless_Parameters}
		\nonumber	&u=\frac{T-T_0}{T_0},\hspace{3mm}q^*=\frac{q}{\rho C_p\left |v  \right |T_0},\hspace{3mm}t^*=\left [ \Gamma(1+\alpha) \right ]^{1/\alpha}\frac{t}{\tau_{q}},\\
		& x^*=\frac{x}{L_c},\hspace{3mm}\textit{Kn}=\frac{\Lambda}{L_c},\hspace{3mm}B=\frac{\tau_t}{\tau_{q}}.
	\end{align}
where, $\Lambda$, $L_c$, $T$, $\tau_q$ and $v$ are, respectively, the phonon mean free-path, the characteristic length, the reference temperature, the heat flux phase lag for silicon with constant bulk thermal properties, and the energy carriers' average velocity (the sound speed). Also, the heat conduction coefficient of the solid material and the sound velocity are defined as $k=\rho C_p|v|\frac{\Lambda}{3}$ and $|v|=\Lambda/\tau_q$. By substituting the non-dimensional parameters and using the mentioned definitions, the non-dimensional fractional DPL equation is achieved:

	\begin{eqnarray} \label{DFDPL*_eq}
		\nonumber 	\frac{\partial u}{\partial t^*}+_{0}^{C}\textrm{\textit{D}}_{t^*}^{\alpha+1} u &=&\frac{\textit{Kn}^2}{3\left [ \Gamma(1+\alpha) \right ]^{1/\alpha}}\left [ \frac{\partial^2 u}{\partial x^{*2}}+B^{\alpha}{}_{0}^{C}\textrm{\textit{D}}_{t^*}^\alpha \left (\frac{\partial^2 u}{\partial x^{*2}}\right)\right ]\\
		&+&F(x^*,t^*).
	\end{eqnarray}

	It is easier to ignore the * sign and write the above equation in the following form:
	\begin{align}
		 \label{DFDPL_eq}
		\nonumber u_t(x,t)+_{0}^{C}\textrm{\textit{D}}_t^{\alpha+1} u(x,t)&=\frac{\textit{Kn}^2}{3\left [ \Gamma(1+\alpha) \right ]^{1/\alpha}}\left ( u+B^{\alpha}{}_{0}^{C}\textrm{\textit{D}}_t^\alpha u \right )_{xx}(x,t) \nonumber\\
		&+F(x,t),\hspace{3mm} 0\leqslant x\leqslant L,\hspace{3mm} 0< t\leqslant \bar{T}.
	\end{align}
	Also, $F(x,T)$ is defined as,
	\begin{equation} \label{DSource_Term}
		F(x,t)=\frac{\tau_q}{\rho C_pT_0\left [ \Gamma(1+\alpha) \right ]^{1/\alpha}}f(x,t).
	\end{equation}	
	The non-dimensional form of the Eqs. (\ref{Energy_eq}) and (\ref{qT_TSE}) are also presented as:

		\begin{align} \label{DEnergy_eq}
		\nonumber	u_t(x,t)=-\frac{\textit{Kn}}{\left [ \Gamma(1+\alpha) \right ]^{1/\alpha}}q_x(x,t)&+F(x,t), \\
		\nonumber &0\leqslant x\leqslant L,\hspace{3mm} 0< t\leqslant \bar{T}\\
		\nonumber	q(x,t)+_{0}^{C}\textrm{\textit{D}}_t^{\alpha} q(x,t)=-\frac{\textit{Kn}}{3}&\left ( u+B^{\alpha}{}_{0}^{C}\textrm{\textit{D}}_t^\alpha u \right )_{x}(x,t), \\
		&0\leqslant x\leqslant L,\hspace{3mm} 0< t\leqslant \bar{T}.
	\end{align}	

	\subsection{Initial and boundary conditions}
	Using the boundary conditions considering no temperature jump, the phonon scattering effects are ignored. Hence, the numerical solutions obtained from the DPL model, especially near the boundaries, are not accurate and are different from the ones calculated using the Boltzmann equation. Actually, the important observed phenomenon dealing with the heat transfer in micro and nano dimensions, is the creation of temperature jump at the boundaries, and also its increase with decreasing the system size. This phenomenon can be modeled, using the mixed type boundary condition, $u_s-u_w=-\lambda\textit{Kn}\frac{\partial u}{\partial \hat{n}}|_\Omega$, at the boundaries of the solution field. In this relation, $u_s$, $u_w$, $n$, $\Omega$ are, subsequently, the jumped temperature at the boundary, the boundary temperature, the unit vector normal to the boundary directed outward, and all the boundaries of the solution domain. Also, $\lambda$ is a constant value, which should be determined. Using this boundary condition along the DPL model, makes the simulation of the temperature jump at the boundary more probable. The values of two unknown parameters of $B$ and $\lambda$ are determined such that the solution obtained from the DPL model matches that of found from the Boltzmann equation, well. For a one-dimensional geometry, the temperature jump boundary conditions at the top and bottom boundaries are correspondingly established as:

	\begin{align}
		\nonumber	&-\lambda\textit{Kn}u_x(0,t)+u(0,t)=\phi_1(t),\hspace{3mm} 0< t\leqslant \bar{T}\\
		\label{Temperature_Jump_RBC}
		&\lambda\textit{Kn}u_x(L,t)+u(L,t)=\phi_2(t),\hspace{3mm} 0< t\leqslant \bar{T}.
	\end{align}

	Moreover, the initial conditions are taken to be:
	\begin{equation}
		\label{ICS}
		u(x,0)=\psi_1(x),\hspace{3mm} u_t(x,0)=\psi_2(x),\hspace{3mm} 0\leqslant x\leqslant L.
	\end{equation}

\section{Solution technique}
\label{Solution technique}
In this section, we will discuss how one can discretize the one-dimensional fractional DPL equation with temperature jump boundary condition. So, the spatial interval $[0,L]$ is discretized into $M$ subintervals, and also the temporal interval $[0,\bar{T}]$ is divided to $K$ intervals:
\begin{align}
	\nonumber
		\label{Temporal_Step}
	&h=\frac{L}{M},\hspace{3mm} x_i=ih,\hspace{3mm} 0\leqslant i\leqslant M\\,
	&\tau=\frac{\bar{T}}{K},\hspace{3mm} t_k=k\tau,\hspace{3mm} 0\leqslant k\leqslant K.
\end{align}

In this relations, $h$ and $\tau$ are space and time steps. Further, as a mean to discretize the space and time derivatives, the following central finite difference operators are utilized:
\begin{align}
	\nonumber
	\label{dt_operator}
	&\delta_xu_{i-\frac{1}{2}}^k=\frac{1}{h}\left ( u_i^k-u_{i-1}^k \right),\hspace{3mm} \delta_x^2u_i^k=\frac{1}{h}\left (\delta_xu_{i+\frac{1}{2}}^k-\delta_xu_{i-\frac{1}{2}}^k \right)\\
	&\delta_tu_{i}^{k-\frac{1}{2}}=\frac{1}{\tau}\left (u_{i}^k-u_{i}^{k-1} \right),\hspace{3mm} u_{i}^{k-\frac{1}{2}}=\frac{1}{2}\left (u_{i}^k+u_{i}^{k-1} \right).
\end{align}

On the other hand, for the purpose of Caputo fractional derivative operator assessment, the $L-1$ approximation is employed. For $0<\alpha<1$, this approximation leads to \cite{ji2018numerical2}:

\begin{align}
	&\textit{D}_t^{\alpha}u_i^k=\frac{\tau^{-\alpha}}{\Gamma(2-\alpha)}\left [ a_0u_i^k-\sum_{n=1}^{k-1}\left ( a_{k-n-1}-a_{k-n} \right )u_i^n-a_{k-1}u_i^0 \right ] \nonumber\\	&\textit{D}_t^{\alpha+1}u_i^k=\frac{\tau^{-\alpha}}{\Gamma(2-\alpha)}\left [ a_0\delta_tu_i^k-\sum_{n=1}^{k-1}\left ( a_{k-n-1}-a_{k-n} \right )\delta_tu_i^n-a_{k-1}\left (u_t  \right )_i^0 \right ] 		\label{L1_approximation_2}
\end{align}

The weighting factors are calculated via $a_l=(l+1)^{1-\alpha}-l^{1-\alpha},\hspace{3mm} l\geqslant0$. In similarity to the Eqs. (\ref{L1_approximation_2}), in order to approximate the Caputo fractional derivative in $k-1/2$ time-step, the following relations are hold:
\begin{eqnarray*}
	\nonumber \textit{D}_t^{\alpha}u_i^{k-\frac{1}{2}}&=&\frac{1}{2}\left (\textit{D}_t^{\alpha}u_i^{k}+\textit{D}_t^{\alpha}u_i^{k-1}\right ) \\
	\nonumber &=&\frac{\tau^{-\alpha}}{\Gamma(2-\alpha)}\left [a_0u_i^{k-\frac{1}{2}}-\sum_{n=1}^{k-1}\left ( a_{k-n-1}-a_{k-n} \right )u_i^{n-\frac{1}{2}}-a_{k-1}u_i^0 \right ] \\
	\nonumber \textit{D}_t^{\alpha+1}u_i^{k-\frac{1}{2}}&=&\frac{1}{2}\left ( \textit{D}_t^{\alpha+1}u_i^{k}+\textit{D}_t^{\alpha+1}u_i^{k-1}\right ) \\
	&=&\frac{\tau^{-\alpha}}{\Gamma(2-\alpha)}\Biggl[ a_0\delta_tu_i^{k-\frac{1}{2}}-\sum_{n=1}^{k-1}\left ( a_{k-n-1}-a_{k-n} \right )\delta_tu_i^{n-\frac{1}{2}}-a_{k-1}\left (u_t  \right )_i^0 \Biggr] 	
\end{eqnarray*}

Furthermore, the second-order derivative of an arbitrary function like $g(x)$ considering $g(x) \in C[x,x_M]$ is approximated contemplating the following equation,

\begin{align}
	&{g}''_i=\frac{1}{h^2}\Bigl (g_{i+1}-2g_i+g_{i-1} \Bigr)=\delta_x^2g_i 	\nonumber\\
	&{g}''_M=\frac{2}{h}\Bigl({g}'_M-\frac{g_M-g_{M-1}}{h}\Bigr)+\frac{h}{3}{g}'''_M 	\label{g(x_M)}	
 \end{align}

Besides, for an arbitrary function of $f(t)$ with the assumptions $f(t) \in C[t_{k-1},t_k]$ and $t_{k-\frac{1}{2}}$, one has,
\begin{equation}
	\label{f(t_k)}
	\frac{1}{2}\Bigl({f}'(t_{k-1})+{f}'(t_{k})\Bigr)\simeq \frac{f(t_{k})-f(t_{k-1})}{\tau}=\delta_t f^{k-\frac{1}{2}}.
\end{equation}

Considering  Eq. (\ref{DFDPL_eq}) in i-th node and using  Eq. (\ref{g(x_M)}), the fractional DPL model for internal nodes are found as,
\begin{align}
	\label{FDPL_i}
	\frac{\textit{d} }{\textit{d} t}u_i(t)+_{0}^{C}\textrm{\textit{D}}_t^{\alpha+1} u_i(t)&=\frac{\textit{Kn}^2}{3\left [ \Gamma(1+\alpha) \right ]^{1/\alpha}}\delta_x^2\left ( u_i(t)+B^{\alpha}{}_{0}^{C}\textrm{\textit{D}}_t^\alpha u_i(t) \right ) \nonumber\\
\end{align}

In the following, considering the above equation in k-th and (k-1)-th time steps ($t=t_k$ and $t=t_{k-1}$), and taking the average of the two obtained equations and also using Eq. (\ref{f(t_k)}), the finite difference scheme for internal nodes is achieved:
\begin{align}
	\label{FDPL_FD_i}
	\delta_tu_i^{k-\frac{1}{2}}+_{0}^{C}\textrm{\textit{D}}_t^{\alpha+1} u_i^{k-\frac{1}{2}}&=\frac{\textit{Kn}^2}{3\left [ \Gamma(1+\alpha) \right ]^{1/\alpha}}\delta_x^2\left ( u_i^{k-\frac{1}{2}}+B^{\alpha}{}_{0}^{C}\textrm{\textit{D}}_t^\alpha u_i^{k-\frac{1}{2}} \right ) \nonumber\\
	&+F_i^{k-\frac{1}{2}},\hspace{3mm} 1\leqslant i\leqslant M-1,\hspace{3mm} 1\leqslant k\leqslant K.
\end{align}

Taking the x-derivative of two sides of the above equation and applying $x \rightarrow x_0$, the following relation is obtained:
\begin{align}
	\label{FDPL_Dx_0}
	\frac{\textit{Kn}^2}{3\left [ \Gamma(1+\alpha) \right ]^{1/\alpha}}\left ( u+B^{\alpha}{}_{0}^{C}\textrm{\textit{D}}_t^\alpha u \right )_{xxx}(x_0,t)&=u_{xt}(x_0,t)+_{0}^{C}\textrm{\textit{D}}_t^{\alpha+1 }u_x(x_0,t) \nonumber\\
	&-F_x(x_0,t).
\end{align}

Also, contemplating Eq. (\ref{DFDPL_eq}) in the top boundary node $(x,t)$ and using Eqs. (\ref{FDPL_Dx_0}), (\ref{g(x_M)}), and (\ref{Temperature_Jump_RBC}), the fractional DPL model for the top boundary node is found:
\begin{align}
	\label{FDPL_0}
	\left ( 1+\frac{h}{3\lambda\textit{Kn}} \right )\left ( \frac{\textit{d} }{\textit{d} t}u_0(t)+_{0}^{C}\textrm{\textit{D}}_t^{\alpha+1} u_0(t) \right )&=\frac{2\textit{Kn}^2}{3h\left [ \Gamma(1+\alpha) \right ]^{1/\alpha}}\delta_x\left ( u_{\frac{1}{2}}(t)+B^{\alpha}{}_{0}^{C}\textrm{\textit{D}}_t^\alpha u_{\frac{1}{2}}(t) \right ) \nonumber\\
	&-\frac{2\textit{Kn}}{3h\lambda\left [ \Gamma(1+\alpha) \right ]^{1/\alpha}}\left ( u_0(t)+B^{\alpha}{}_{0}^{C}\textrm{\textit{D}}_t^\alpha u_0(t) \right )+W_0(t).
\end{align}
Here, $W_0(t)$ is defined as,
\begin{align}
	\label{W_0(t)}
	W_0(t)&=\frac{2\textit{Kn}}{3h\lambda\left [ \Gamma(1+\alpha) \right ]^{1/\alpha}}\left ( \phi_1(t)+B^{\alpha}{}_{0}^{C}\textrm{\textit{D}}_t^\alpha \phi_1(t) \right ) \nonumber\\
	&+\frac{h}{3\lambda\textit{Kn}}\left ( \frac{\textit{d} }{\textit{d} t}\phi_1(t)+_{0}^{C}\textrm{\textit{D}}_t^{\alpha+1} \phi_1(t) \right )+\frac{h}{3}(F_x)_0(t)+F_0(t).
\end{align}

In the next step, with consideration of Eq. (\ref{FDPL_0}) in k-th time-step ($t=t_k$) and (k-1)-th time-step ($t=t_{k-1}$), and averaging the two obtained equation, and also using Eq. (\ref{f(t_k)}), the finite difference scheme for the top boundary node is obtained,
\begin{align}
	\label{FDPL_FD_0}
	&\left ( 1+\frac{h}{3\lambda\textit{Kn}} \right )\left (\delta_tu_0^{k-\frac{1}{2}}+_{0}^{C}\textrm{\textit{D}}_t^{\alpha+1} u_0^{k-\frac{1}{2}} \right )=\frac{2\textit{Kn}^2}{3h\left [ \Gamma(1+\alpha) \right ]^{1/\alpha}}\delta_x\left ( u_{\frac{1}{2}}^{k-\frac{1}{2}}+B^{\alpha}{}_{0}^{C}\textrm{\textit{D}}_t^\alpha u_{\frac{1}{2}}^{k-\frac{1}{2}} \right ) \nonumber\\
	&-\frac{2\textit{Kn}}{3h\lambda\left [ \Gamma(1+\alpha) \right ]^{1/\alpha}}\left ( u_0^{k-\frac{1}{2}}+B^{\alpha}{}_{0}^{C}\textrm{\textit{D}}_t^\alpha u_0^{k-\frac{1}{2}} \right )+W_0^{k-\frac{1}{2}},\hspace{3mm} 1\leqslant k\leqslant K.
\end{align}
In the above relation, $W^{k-\frac{1}{2}}_{0}$ is,
\begin{align}
	\label{W_0(k-1/2)}
	W_0^{k-\frac{1}{2}}&=\frac{2\textit{Kn}}{3h\lambda\left [ \Gamma(1+\alpha) \right ]^{1/\alpha}}\left ( \phi_1^{k-\frac{1}{2}}+B^{\alpha}{}_{0}^{C}\textrm{\textit{D}}_t^\alpha \phi_1^{k-\frac{1}{2}} \right ) \nonumber\\
	&+\frac{h}{3\lambda\textit{Kn}}\left ( \frac{\textit{d} }{\textit{d} t}\phi_1^{k-\frac{1}{2}}+_{0}^{C}\textrm{\textit{D}}_t^{\alpha+1} \phi_1^{k-\frac{1}{2}} \right )+\frac{h}{3}(F_x)_0^{k-\frac{1}{2}}+F_0^{k-\frac{1}{2}}.
\end{align}
Similarly, repeating the above discretization method, one acquires the finite difference scheme for the bottom boundary node,
\begin{align}
	\label{FDPL_FD_M}
	&\left ( 1+\frac{h}{3\lambda\textit{Kn}} \right )\left (\delta_tu_M^{k-\frac{1}{2}}+_{0}^{C}\textrm{\textit{D}}_t^{\alpha+1} u_M^{k-\frac{1}{2}} \right ) \nonumber\\
	&=-\frac{2\textit{Kn}^2}{3h\left [ \Gamma(1+\alpha) \right ]^{1/\alpha}}\delta_x\left ( u_{M-\frac{1}{2}}^{k-\frac{1}{2}}+B^{\alpha}{}_{0}^{C}\textrm{\textit{D}}_t^\alpha u_{M-\frac{1}{2}}^{k-\frac{1}{2}} \right ) \nonumber\\
	&-\frac{2\textit{Kn}}{3h\lambda\left [ \Gamma(1+\alpha) \right ]^{1/\alpha}}\left ( u_M^{k-\frac{1}{2}}+B^{\alpha}{}_{0}^{C}\textrm{\textit{D}}_t^\alpha u_M^{k-\frac{1}{2}} \right )+W_M^{k-\frac{1}{2}},\hspace{3mm} 1\leqslant k\leqslant K.
\end{align}
Where, $W^{k-\frac{1}{2}}_{M}$ is calculated through,
\begin{align}
	\label{W_M(k-1/2)}
	W_M^{k-\frac{1}{2}}&=\frac{2\textit{Kn}}{3h\lambda\left [ \Gamma(1+\alpha) \right ]^{1/\alpha}}\left ( \phi_2^{k-\frac{1}{2}}+B^{\alpha}{}_{0}^{C}\textrm{\textit{D}}_t^\alpha \phi_2^{k-\frac{1}{2}} \right ) \nonumber\\
	&+\frac{h}{3\lambda\textit{Kn}}\left ( \frac{\textit{d} }{\textit{d} t}\phi_2^{k-\frac{1}{2}}+_{0}^{C}\textrm{\textit{D}}_t^{\alpha+1} \phi_2^{k-\frac{1}{2}} \right )-\frac{h}{3}(F_x)_M^{k-\frac{1}{2}}+F_M^{k-\frac{1}{2}}.
\end{align}
The presented finite difference scheme for the fractional DPL model with temperature jump boundary condition, is unconditionally stable and convergent. The order of convergence based on the infinity norm, is the order of $2-\alpha$ and $2$, respectively, on time and space \cite{ji2018numerical2}. When the Caputo fractional derivative operator approximation is placed in Eqs. (\ref{FDPL_FD_i}), (\ref{FDPL_FD_0}), and (\ref{FDPL_FD_M}), the finite difference discretized form of the fractional DPL model with temperature jump boundary condition is achieved,
\begin{align}
	\label{FDPL_FD_i=0}
	&\left ( d_1c_1+2d_2c_2+d_3c_3 \right )u_0^k+\left ( -2d_2c_2 \right )u_1^k=\left ( d_1c_1-2d_2c_2-d_3c_3 \right )u_0^{k-1} \nonumber\\
	&+\left ( 2d_2c_2 \right )u_1^{k-1}+c_1\mu\Biggl[ \sum_{n=1}^{k-1}\left ( a_{k-n-1}-a_{k-n} \right )\left ( \frac{u_0^n-u_0^{n-1}}{\tau} \right )+a_{k-1}(\psi_2)_0\Biggr] \nonumber\\
	&-\frac{2}{h}B^\alpha\mu c_2\Biggl[ \sum_{n=1}^{k-1}\left ( a_{k-n-1}-a_{k-n} \right )\left ( \frac{u_1^n+u_1^{n-1}-u_0^n-u_0^{n-1}}{2h} \right )+a_{k-1}\left ( \frac{u_1^0-u_0^0}{h} \right ) \Biggr] \nonumber\\
	&+\frac{2}{h\lambda}B^\alpha\mu c_3\Biggl[ \sum_{n=1}^{k-1}\left ( a_{k-n-1}-a_{k-n} \right )\left ( \frac{u_0^n+u_0^{n-1}}{2} \right )+a_{k-1}u_0^0 \Biggr]+\frac{2}{h\lambda}c_3\Biggl\{\frac{\phi_1^k+\phi_1^{k-1}}{2} \nonumber\\
	&+B^\alpha\mu \Biggl[ \frac{a_0}{2}\left ( \phi_1^k+\phi_1^{k-1} \right )-\sum_{n=1}^{k-1}\left ( a_{k-n-1}-a_{k-n} \right )\left ( \frac{\phi_1^n+\phi_1^{n-1}}{2} \right )-a_{k-1}\phi_1^0 \Biggr] \Biggr\} \nonumber\\
	&+\frac{h}{3\lambda\textit{Kn}} \Biggl\{\frac{\phi_1^k-\phi_1^{k-1}}{\tau}+\mu \Biggl[ \frac{a_0}{\tau}\left ( \phi_1^k-\phi_1^{k-1} \right )-\sum_{n=1}^{k-1}\left ( a_{k-n-1}-a_{k-n} \right )\left ( \frac{\phi_1^n-\phi_1^{n-1}}{\tau} \right ) \nonumber\\
	&-a_{k-1}(\phi_1)_t^0 \Biggr] \Biggr\}+\frac{h}{3}\left ( \frac{(F_x)_0^k+(F_x)_0^{k-1}}{2}\right )+\frac{F_0^k+F_0^{k-1}}{2},\hspace{3mm} 1\leqslant k\leqslant K.
\end{align}
\begin{align}
	\label{FDPL_FD_0<i<M}
	&\left ( -d_2c_2 \right )u_{i+1}^k\left ( d_1+2d_2c_2 \right )u_i^k+\left ( -d_2c_2 \right )u_{i-1}^k=\left ( d_1-2d_2c_2 \right )u_i^{k-1} \nonumber\\
	&+d_2c_2 \left ( u_{i+1}^{k-1}+u_{i-1}^{k-1} \right )+\mu\Biggl[ \sum_{n=1}^{k-1}\left ( a_{k-n-1}-a_{k-n} \right )\left ( \frac{u_i^n-u_i^{n-1}}{\tau} \right )+a_{k-1}(\psi_2)_i\Biggr] \nonumber\\
	&-B^\alpha\mu c_2 \Biggl[ \sum_{n=1}^{k-1}\left ( a_{k-n-1}-a_{k-n} \right )\left ( \frac{u_{i+1}^n+u_{i+1}^{n-1}-2u_{i}^n-2u_{i}^{n-1}+u_{i-1}^n+u_{i-1}^{n-1}}{2h^2} \right ) \nonumber\\
	&+a_{k-1}\left ( \frac{u_{i+1}^0-2u_{i}^0+u_{i-1}^0}{h^2} \right ) \Biggr] \Biggr\}+\frac{F_i^k+F_i^{k-1}}{2},\hspace{3mm} 1\leqslant i\leqslant M-1,\hspace{3mm} 1\leqslant k\leqslant K \\
	\label{FDPL_FD_i=M}
	&\left ( d_1c_1+2d_2c_2+d_3c_3 \right )u_M^k+\left ( -2d_2c_2 \right )u_{M-1}^k=\left ( d_1c_1-2d_2c_2-d_3c_3 \right )u_M^{k-1} \nonumber\\
	&+\left ( 2d_2c_2 \right )u_{M-1}^{k-1}+c_1\mu\Biggl[ \sum_{n=1}^{k-1}\left ( a_{k-n-1}-a_{k-n} \right )\left ( \frac{u_M^n-u_M^{n-1}}{\tau} \right )+a_{k-1}(\psi_2)_M\Biggr] \nonumber\\
	&+\frac{2}{h}B^\alpha\mu c_2 \Biggl[ \sum_{n=1}^{k-1}\left ( a_{k-n-1}-a_{k-n} \right )\left ( \frac{u_M^n+u_M^{n-1}-u_{M-1}^n-u_{M-1}^{n-1}}{2h} \right )+a_{k-1}\left ( \frac{u_M^0-u_{M-1}^0}{h} \right ) \Biggr] \nonumber\\
	&+\frac{2}{h\lambda}B^\alpha\mu c_3\Biggl[ \sum_{n=1}^{k-1}\left ( a_{k-n-1}-a_{k-n} \right )\left ( \frac{u_M^n+u_M^{n-1}}{2} \right )+a_{k-1}u_M^0 \Biggr]+\frac{2}{h\lambda}c_3\Biggl\{\frac{\phi_2^k+\phi_2^{k-1}}{2} \nonumber\\
	&+B^\alpha\mu \Biggl[ \frac{a_0}{2}\left ( \phi_2^k+\phi_2^{k-1} \right )-\sum_{n=1}^{k-1}\left ( a_{k-n-1}-a_{k-n} \right )\left ( \frac{\phi_2^n+\phi_2^{n-1}}{2} \right )-a_{k-1}\phi_2^0 \Biggr] \Biggr\} \nonumber\\
	&+\frac{h}{3\lambda\textit{Kn}} \Biggl\{\frac{\phi_2^k-\phi_2^{k-1}}{\tau}+\mu \Biggl[ \frac{a_0}{\tau}\left ( \phi_2^k-\phi_2^{k-1} \right )-\sum_{n=1}^{k-1}\left ( a_{k-n-1}-a_{k-n} \right )\left ( \frac{\phi_2^n-\phi_2^{n-1}}{\tau} \right ) \nonumber\\
	&-a_{k-1}(\phi_2)_t^0 \Biggr] \Biggr\}-\frac{h}{3}\left ( \frac{(F_x)_M^k+(F_x)_M^{k-1}}{2}\right )+\frac{F_M^k+F_M^{k-1}}{2},\hspace{3mm} 1\leqslant k\leqslant K
\end{align}
In the relations above, $\mu$, $d_1$, $d_2$, $d_3$, $c_1$, $c_2$ and $c_3$ are defined as,
\begin{align}
	\label{Parameters}
	&\mu=\frac{\tau^{-\alpha}}{\Gamma(2-\alpha)},\hspace{3mm} d_1=\frac{\mu a_0+1}{\tau},\hspace{3mm} d_2=\frac{B^{\alpha}\mu a_0+1}{2h^2},\hspace{3mm} d_3=\frac{B^{\alpha}\mu a_0+1}{h\lambda},\hspace{3mm} \nonumber\\ &c_1=1+\frac{h}{3\lambda\textit{Kn}},\hspace{3mm} c_2=\frac{\textit{Kn}^2}{3\left [ \Gamma(1+\alpha) \right ]^{1/\alpha}},\hspace{3mm}c_3=\frac{\textit{Kn}}{3\left [ \Gamma(1+\alpha) \right ]^{1/\alpha}}.
\end{align}
\section{Verification}
\label{verification}
\begin{figure}[htbp]
	\hspace{-1cm}
	\includegraphics[scale=0.35]{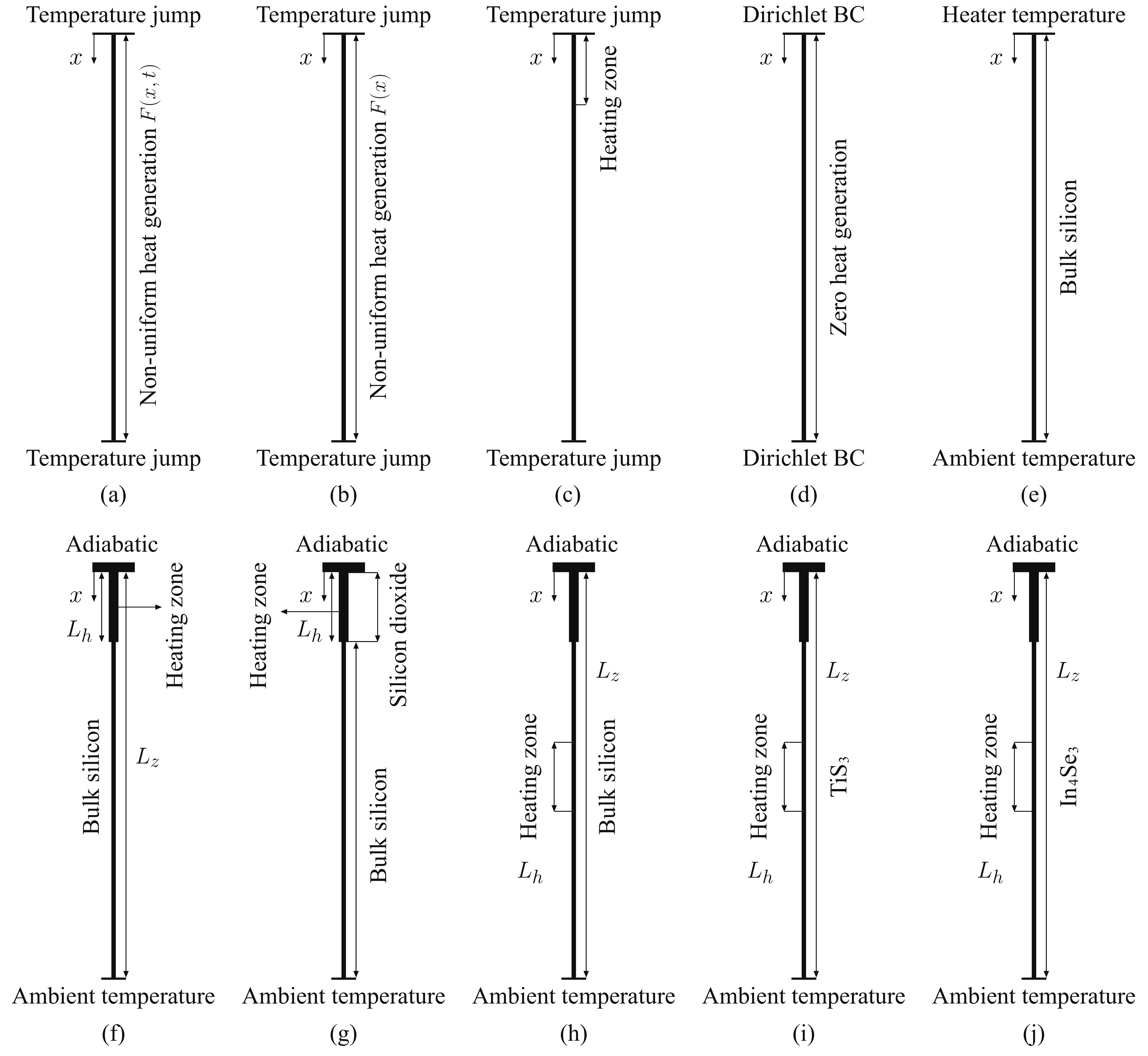}
	\caption{Schematic geometry and corresponding boundary conditions of different modeled transistors, (a) the slab with non-uniform heat generation as a function of time and space, (b) the slab with non-uniform heat generation as a function of space, (c) the silicon slab with uniform heat generation at the upper part, (d) the slab without the source term, (e) the silicone slab without the source term, (f) the silicone MOSFET with uniform heat generation at the upper part, (g) the MOSFET with the buried oxide with uniform heat generation, (h) the silicon MOSFET with uniform heat generation at the middle part, (i) the TiS$_3$ MOSFET with uniform heat generation at the middle part, (j) the In$_4$Se$_3$ MOSFET with uniform heat generation at the middle part.}
	\label{CasesSchematic}
\end{figure}

In this section, first, in order to test the accuracy of the presented finite difference scheme, five different one-dimensional cases with the temperature jump boundary condition and a source heat are investigated. The obtained results including the convergence order and the error in base of the infinity norm, are verified with the available data in \cite{ji2018numerical2} to be sure of accuracy of the derived procedure. Then, the results of heat transport investigation in the newly proposed quasi one-dimensional transistors will be presented. The geometry of cases which are studied in the present work are shown in Fig. \ref{CasesSchematic}. Also, the verifications performed for different cases and the results obtained from various states of MOSFET transistors are all gathered in Table. \ref{results_table}.
\begin{table}[htbp]
	\caption{Details of different cases for validation and results.}
	\label{results_table}
	\begin{center}
		\begin{small}
			\begin{tabular}{r@{\hspace{0.5cm}}c@{\hspace{0.5cm}}c@{\hspace{0.5cm}}c@{\hspace{0.5cm}}c}
				\hline
				\hline
				Verification cases&&&&\\
				\hline
				Type of transistor&Channel material&Heat generation&Buried oxide&bulk/film properties\\
				\hline\hline
				Case one&-& Entire transistor& No & Bulk \\
				\hline
				Case two&-& Entire transistor& No & Bulk \\
				\hline
				Case three&Si& The beginning & No & Bulk \\
				\hline
				Case four&-& No & No & Bulk \\
				\hline
				Case five&Si& No & No & Bulk \\
				\hline
				\hline
				Final results &&&&\\
				\hline
				Type of transistor&Channel material&Heat generation&Buried oxide&bulk/film properties\\\hline\hline
				Case I(a)&Si& The beginning  & No & Bulk \\
				\hline
				Case I(b)&Si&  The beginning & No & Film \\
				\hline
				Case II(a)&Si&  The beginning & Yes & Bulk \\
				\hline
				Case II(b)&Si&  The beginning & Yes & Film \\
				\hline
				Case III&Si& The middle & No & Film \\
				\hline
				Case IV&$TiS_{3}$& The middle & No & Film \\
				\hline
				Case V& $In_{4}Se_{3}$& The middle & No & Film \\
				\hline\hline
			\end{tabular}
		\end{small}
	\end{center}
\end{table}
\subsection{Mesh/time-step size independence tests}
The results of the time/step size independence test for the first geometry, Fig. \ref{CasesSchematic}(a), are demonstrated in Table. \ref{GIS}. The results are shown for three different temperatures and also for five values of position. It is found that the results for the meshes of (M=2000, K=3000) and (M=2200, K=3200) are similar. Hence, to reduce the computational cost, the mesh, M=2000, K=3000, is used to proceed the modeling.
\begin{table}[htbp]
	\caption{Results of the mesh size and the time-step size independence tests at $t=10\hspace{1mm}\textup{ps}.$}
	\label{GIS}
	\begin{center}
		\begin{small}
			\begin{tabular}{ccccccc}
				\hline
				$(M,K)$&$x = 0 \hspace{1mm}\textup{nm}$&$x = 10 \hspace{1mm}\textup{nm}$&$x = 20 \hspace{1mm}\textup{nm}$&$x = 30 \hspace{1mm}\textup{nm}$&$x = 40 \hspace{1mm}\textup{nm}$&$x = 50 \hspace{1mm}\textup{nm}$\\
				\hline
				$(1000,2000)$&$307.2538	
				$&$304.5218	
				$&$300.9154	
				$&$300.0286	
				$&$300.0000	
				$&$300.0000	
				$\\
				\hline
				$(1200,2200)$&$307.2537	
				$&$304.5195	
				$&$300.9144	
				$&$300.0285	
				$&$300.0000	
				$&$300.0000	
				$\\
				\hline
				$(1400,2400)$&$307.2536	
				$&$304.5179	
				$&$300.9137	
				$&$300.0284	
				$&$300.0000	
				$&$300.0000	
				$\\
				\hline
				$(1600,2600)$&$307.2536	
				$&$304.5166	
				$&$300.9132	
				$&$300.0284	
				$&$300.0000	
				$&$300.0000	
				$\\
				\hline
				$(1800,2800)$&$307.2536	
				$&$304.5157	
				$&$300.9127	
				$&$300.0283	
				$&$300.0000	
				$&$300.0000	
				$\\
				\hline
				$(2000,3000)$&$307.2535	
				$&$304.5149	
				$&$300.9124	
				$&$300.0283	
				$&$300.0000	
				$&$300.0000	
				$\\
				\hline
				$(2200,3200)$&$307.2535	
				$&$304.5142	
				$&$300.9121	
				$&$300.0283	
				$&$300.0000	
				$&$300.0000	
				$\\
				\hline
			\end{tabular}
		\end{small}
	\end{center}
\end{table}
\subsection{Verification: Case one}
\begin{figure}[htbp]
	\begin{center}
		\includegraphics[scale=0.5]{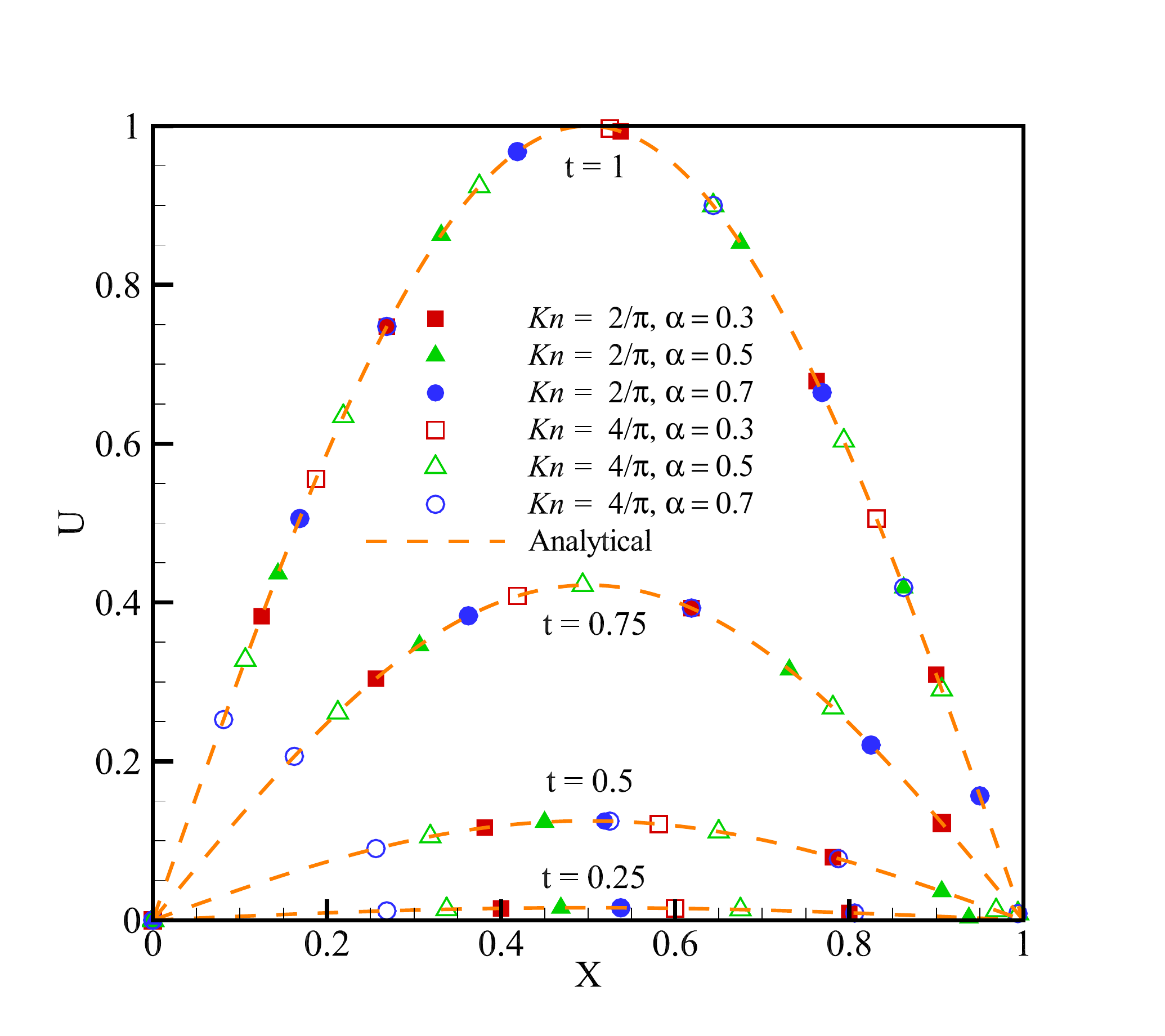}
	\end{center}
	\caption{Comparison of analytical and numerical results of Ji \emph{et al.}~\cite{ji2018numerical2} considering the case 1 for different values of $\alpha$, h=0.00625, and $\tau$=0.001.}
	\label{Fig.3.2}
\end{figure}

Here, the one-dimensional fractional DPL model for the case containing the temperature jump boundary condition and the non-homogeneous heat source, which is presented in Fig. \ref{CasesSchematic}(a), is studied. The governing equation, the boundary and initial conditions for this case are respectively as,
\begin{align} \label{DFDPL_eq_P1}
	u_t(x,t)+_{0}^{C}\textrm{\textit{D}}_t^{\alpha+1} u(x,t)&=\frac{\textit{Kn}^2}{3\left [ \Gamma(1+\alpha) \right ]^{1/\alpha}}\left ( u+B^{\alpha}{}_{0}^{C}\textrm{\textit{D}}_t^\alpha u \right )_{xx}(x,t) \nonumber\\
	&+F(x,t),\hspace{3mm} 0\leqslant x\leqslant 1,\hspace{3mm} 0< t\leqslant 1
\end{align}
\begin{align}
	\nonumber	&-\lambda\textit{Kn}u_x(0,t)+u(0,t)=-\pi\lambda\textit{Kn}t^3,\hspace{3mm} 0< t\leqslant 1\\
	\nonumber 	&\lambda\textit{Kn}u_x(1,t)+u(1,t)=-\pi\lambda\textit{Kn}t^3,\hspace{3mm} 0< t\leqslant 1\\
	\label{ICS_P1}
	&u(x,0)=0,\hspace{3mm} u_t(x,0)=0,\hspace{3mm} 0\leqslant x\leqslant 1
\end{align}
The considered source heat $F(x,t)$ is,
\begin{align}
	\label{Source_Term_P1}
	F(x,t)=\Biggl[ 3t^2+\frac{6}{\Gamma(3-\alpha)}t^{2-\alpha}+\frac{\pi^2\textit{Kn}^2}{3\left [ \Gamma(1+\alpha) \right ]^{1/\alpha}}\left ( t^3+\frac{6B^\alpha}{\Gamma(4-\alpha)}t^{3-\alpha} \right )\Biggr]\sin(\pi x)
\end{align}
Also, the analytical solution is found to be $U(x,t)=t^3\sin(\pi x)$ \cite{ji2018numerical2}. The following relations are used to calculate the numerical errors in base of the infinity norm and the convergence order,
\begin{align}
	\nonumber 	&E_{\infty}(h)=\left \| U^K-u^K \right \|_{\infty}\\
	\label{Rate1}
	&Rate_1=\log_2\left ( \frac{E_{\infty}(2h)}{E_{\infty}(h)} \right ).
\end{align}
In order to find the convergence order, the time interval is divided into $K=\bar{T}(\frac{M}{L}^{\frac{2}{2-\alpha}})$ subintervals. In Tab. \ref{Tabel_P1_1}, our obtained results using the infinity norm and the convergence order are compared with those calculated by Ji \emph{et al.}\cite{ji2018numerical2} for $Kn=\frac{2}{\pi}<1$, $\lambda=\frac{1}{5}$, $B=\frac{1}{2}$, and different values of $\alpha$.
\begin{table}[htbp]
	\caption{The infinity norm and the convergence order for $Kn=\frac{2}{\pi}<1$.}
	\label{Tabel_P1_1}
	\begin{center}
		\begin{small}
			\begin{tabular}{llcllcll}
				\hline
				\multicolumn{2}{c}{Ji et al.~\cite{ji2018numerical2}}&&\multicolumn{2}{c}{Present results} &&$\textit{h}$&$\alpha$\\ [0.5ex]
				\cline{1-2}
				\cline{4-5}
				\multicolumn{2}{l}{$E_{\infty}(h)\hspace{15mm}Rate_1$}&&\multicolumn{2}{l}{$E_{\infty}(h)\hspace{15mm}Rate_1$}&&&\\
				\hline
				&$1.637\times 10^{-3}$&&&$1.840\times 10^{-3}$&&$0.05$&$0.3$\\
				$2.012$&$4.057\times 10^{-4}$&&$2.001$&$4.597\times 10^{-4}$&&$0.025$&\\
				$1.997$&$1.016\times 10^{-4}$&&$1.998$&$1.151\times 10^{-4}$&&$0.0125$&\\
				$2.010$&$2.522\times 10^{-5}$&&$1.994$&$2.890\times 10^{-5}$&&$0.00625$&\\
				&$1.982\times 10^{-3}$&&&$2.112\times 10^{-3}$&&$0.05$&$0.5$\\
				$2.015$&$4.902\times 10^{-4}$&&$1.981$&$5.349\times 10^{-4}$&&$0.025$&\\
				$2.002$&$1.224\times 10^{-4}$&&$1.999$&$1.338\times 10^{-4}$&&$0.0125$&\\
				$2.001$&$3.057\times 10^{-5}$&&$1.999$&$3.346\times 10^{-5}$&&$0.00625$&\\
				&$2.380\times 10^{-3}$&&&$2.477\times 10^{-3}$&&$0.05$&$0.7$\\
				$2.009$&$5.913\times 10^{-4}$&&$1.993$&$6.223\times 10^{-4}$&&$0.025$&\\
				$1.998$&$1.480\times 10^{-4}$&&$1.998$&$1.558\times 10^{-4}$&&$0.0125$&\\
				$2.000$&$3.700\times 10^{-5}$&&$2.000$&$3.894\times 10^{-5}$&&$0.00625$&\\
				\hline
			\end{tabular}
		\end{small}
	\end{center}
\end{table}
Also, analogously, the numerical results comparison for the parameters $Kn=\frac{4}{\pi}<1$, $\lambda=\frac{1}{10}$, and $B=\frac{4}{3}$ are displayed in Tab. \ref{Tabel_P1_2}. According to Tabs. \ref{Tabel_P1_1} and \ref{Tabel_P1_2}, as one expects, the convergence order goes to $O(h^2)$, and also as time and space steps decrease, the value of the numerical error based on the infinity norm reduces. As it is shown in Tab. \ref{Tabel_P1_1}, the average relative error between our results and the one reported in \cite{ji2018numerical2}, for the infinity norm error and the convergence error at $Kn=\frac{2}{\pi}<1$ are obtained. The error is calculated using the relative error formula which, here, is the difference between the obtained results and the result reported in \cite{ji2018numerical2}, divided by the available data \cite{ji2018numerical2}. The infinity norm error and the convergence error are obtained, subsequently, as 9 and 0.5 percent. Also, the Tab. \ref{Tabel_P1_2} demonstrates that the obtained errors when $Kn=\frac{4}{\pi}>1$ holds, are less than 20 and 0.5 percent. Moreover, comparison between the analytical results and the numerical data obtained from the finite difference scheme is shown in Fig. \ref{Fig.3.2}. It is seen that the numerical results corresponding to $Kn=\frac{2}{\pi}$ and $Kn=\frac{4}{\pi}$ for $\alpha=0.7$ present acceptable consistency with the analytical results, and the average relative error for both results is less than one percent.
\begin{table}[htbp]
	\caption{The infinity norm and the rate of convergence for $Kn=\frac{4}{\pi}>1$.}
	\label{Tabel_P1_2}
	\begin{center}
		\begin{small}
			\begin{tabular}{llcllcll}
				\hline
				\multicolumn{2}{c}{Ji et al.~\cite{ji2018numerical2}}&&\multicolumn{2}{c}{Present results} &&$\textit{h}$&$\alpha$\\ [0.5ex]
				\cline{1-2}
				\cline{4-5}
				\multicolumn{2}{l}{$E_{\infty}(h)\hspace{15mm}Rate_1$}&&\multicolumn{2}{l}{$E_{\infty}(h)\hspace{15mm}Rate_1$}&&&\\
				\hline
				&$2.498\times 10^{-3}$&&&$3.013\times 10^{-3}$&&$0.05$&$0.3$\\
				$1.985$&$6.309\times 10^{-4}$&&$1.977$&$7.654\times 10^{-4}$&&$0.025$&\\
				$1.981$&$1.598\times 10^{-4}$&&$1.980$&$1.939\times 10^{-4}$&&$0.0125$&\\
				$1.993$&$4.016\times 10^{-5}$&&$1.981$&$4.912\times 10^{-5}$&&$0.00625$&\\
				&$3.087\times 10^{-3}$&&&$3.645\times 10^{-3}$&&$0.05$&$0.5$\\
				$1.997$&$7.732\times 10^{-4}$&&$1.972$&$9.289\times 10^{-4}$&&$0.025$&\\
				$1.993$&$1.943\times 10^{-4}$&&$1.990$&$2.339\times 10^{-4}$&&$0.0125$&\\
				$1.995$&$4.873\times 10^{-5}$&&$1.993$&$5.873\times 10^{-5}$&&$0.00625$&\\
				&$3.685\times 10^{-3}$&&&$4.323\times 10^{-3}$&&$0.05$&$0.7$\\
				$2.002$&$9.200\times 10^{-4}$&&$1.989$&$1.089\times 10^{-3}$&&$0.025$&\\
				$1.997$&$2.306\times 10^{-4}$&&$1.996$&$2.730\times 10^{-4}$&&$0.0125$&\\
				$1.999$&$5.768\times 10^{-5}$&&$1.999$&$6.832\times 10^{-5}$&&$0.00625$&\\
				\hline
			\end{tabular}
		\end{small}
	\end{center}
\end{table}
\subsection{Verification: Case two}
In this section, the case presented in Fig. \ref{CasesSchematic}(b) has been investigated. Due to the lack of analytical solutions, the maximum numerical error and convergence error are calculated via,
\begin{align}
	\nonumber 	&H_{\infty}(h)=\max \left | u_i^K(h,\tau)-u_{2i}^{2K}(\frac{h}{2},\frac{\tau}{2} )\right |,\hspace{3mm} 0\leqslant i\leqslant M\\
	\label{Rate2}
	&Rate_2=\log_2\left ( \frac{H_{\infty}(2h)}{H_{\infty}(h)} \right )
\end{align}
Here, the governing equation and the boundary equations are described respectively, by Eqs. (\ref{DFDPL_eq}), (\ref{Temperature_Jump_RBC}), and (\ref{ICS}). The maximum numerical error and the convergence order for this case when $\bar{T}=1$, $L=1$, $B=0.25$, $\lambda=1$, $\phi_1(t)=0$, $\phi_2(t)=2$, $\psi_1=0$, $\psi_2=0$, $F(x,t)=sin(x)$, and $Kn=10$. The obtained results are dispensed in Tab. \ref{Tabel_P2_3}.
\begin{table}[htbp]
	\caption{Error analysis and the convergence order at $Kn=10$.}
	\label{Tabel_P2_3}
	\begin{center}
		\begin{small}
			\begin{tabular}{llcllcll}
				\hline
				\multicolumn{2}{c}{Ji et al.~\cite{ji2018numerical2}}&&\multicolumn{2}{c}{Present results} &&$\textit{h}$&$\alpha$\\ [0.5ex]
				\cline{1-2}
				\cline{4-5}
				\multicolumn{2}{l}{$H_{\infty}(h)\hspace{15mm}Rate_2$}&&\multicolumn{2}{l}{$H_{\infty}(h)\hspace{15mm}Rate_2$}&&&\\
				\hline
				&$8.262\times 10^{-6}$&&&$5.698\times 10^{-6}$&&$0.05$&$0.3$\\
				$2.225$&$1.767\times 10^{-6}$&&$2.261$&$1.189\times 10^{-6}$&&$0.025$&\\
				$2.110$&$4.093\times 10^{-7}$&&$2.089$&$2.796\times 10^{-7}$&&$0.0125$&\\
				$2.089$&$9.623\times 10^{-8}$&&$2.046$&$6.769\times 10^{-8}$&&$0.00625$&\\
				&$3.429\times 10^{-6}$&&&$3.819\times 10^{-6}$&&$0.05$&$0.5$\\
				$2.326$&$6.837\times 10^{-7}$&&$2.195$&$8.340\times 10^{-7}$&&$0.025$&\\
				$2.279$&$1.409\times 10^{-7}$&&$2.125$&$1.912\times 10^{-7}$&&$0.0125$&\\
				$2.208$&$3.050\times 10^{-8}$&&$2.086$&$4.502\times 10^{-8}$&&$0.00625$&\\
				&$9.569\times 10^{-6}$&&&$6.985\times 10^{-6}$&&$0.05$&$0.7$\\
				$1.938$&$2.498\times 10^{-6}$&&$1.882$&$1.895\times 10^{-6}$&&$0.025$&\\
				$1.954$&$6.447\times 10^{-7}$&&$1.950$&$4.903\times 10^{-7}$&&$0.0125$&\\
				$1.977$&$1.638\times 10^{-7}$&&$1.982$&$1.241\times 10^{-7}$&&$0.00625$&\\
				\hline
			\end{tabular}
		\end{small}
	\end{center}
\end{table}
As the results suggest, as the space and time steps decrease, the value of the maximum numerical error reduces and the convergence order goes to $O(h^2)$.
\subsection{Verification: Case three}
\begin{figure}[htbp]
	\begin{center}
		\includegraphics[scale=0.6]{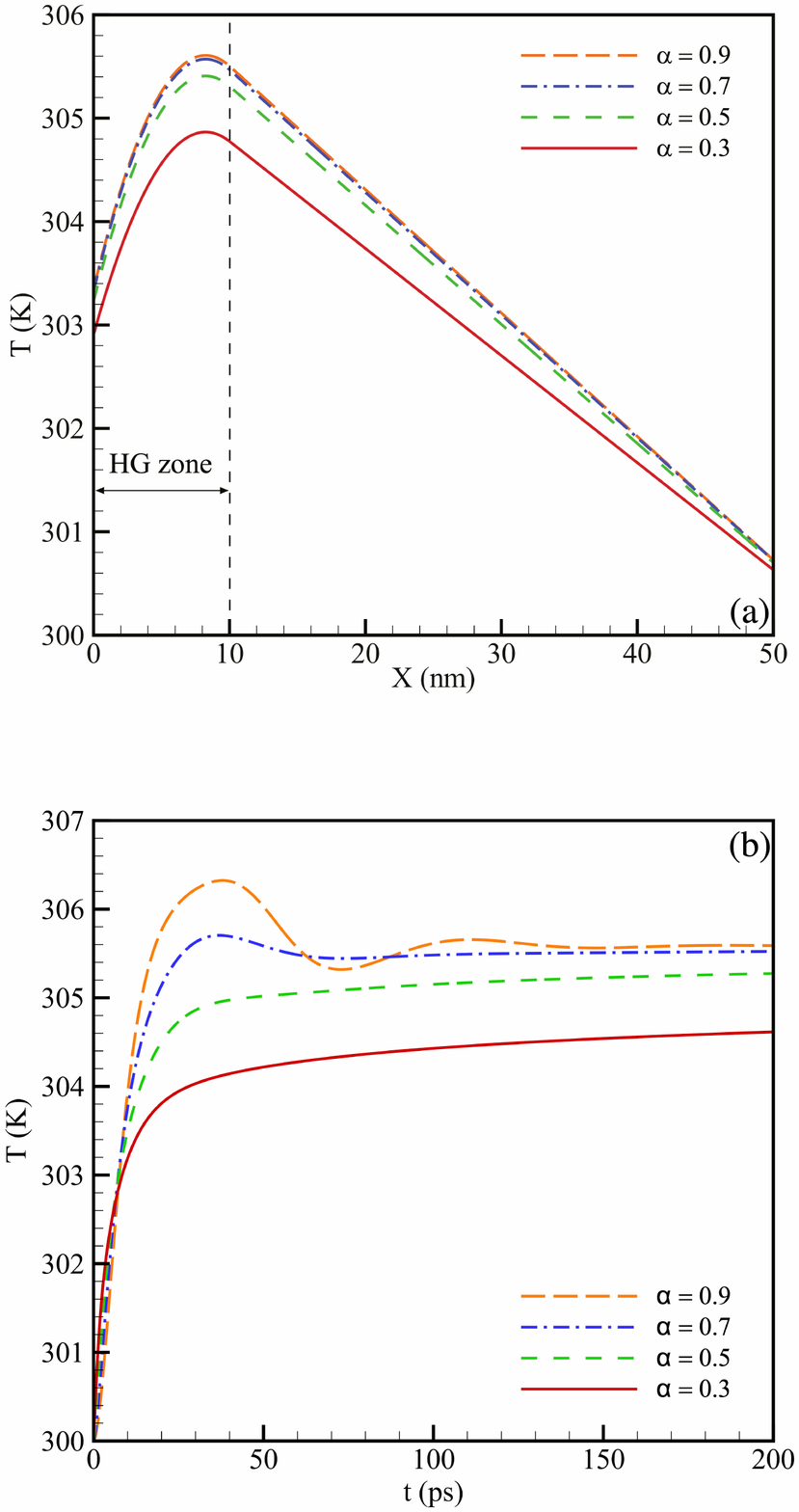}
	\end{center}
	\caption{a) Temperature distribution for the case three of verification at t=200 ps, (b) temporal variation of temperature distribution at X=7.5 nm and Kn=10.}
	\label{case3}
\end{figure}

Here, thermal analysis of a silicon slab with uniform heat generation zone and temperature jump boundary condition like what presented in Fig. \ref{CasesSchematic} has been performed using a fractional DPL model. The thermal properties of silicon are presented in Tab. \ref{Tabel_P3_1}.
\begin{table}[htbp]
	\caption{Thermal properties of silicon~\cite{ghazanfarianshomali2012investigation}}
	\label{Tabel_P3_1}
	\begin{center}
		\begin{small}
			\begin{tabular}{l@{\hspace{1cm}}l@{\hspace{1cm}}l@{\hspace{1cm}}l@{\hspace{1cm}}l@{\hspace{1cm}}l}
				\hline
				$\rho C_p(\hspace{1mm}\textrm{J}\textup{m}^{-3}\textup{K}^{-1})   $&$\tau_q(\textup{ps})$&$L_z(\textup{nm})$&$\Lambda(\textup{nm})$&$L_c(\textup{nm})$&$T_0(\textup{K})$\\
				\hline
				$1.5\times 10^6$&$33.33$&$50$&$100$&$10$&$300$\\
				\hline
			\end{tabular}
		\end{small}
	\end{center}
\end{table}
Again, the governing equation and the boundary equations are presented, respectively, by Eqs. (\ref{DFDPL_eq}), (\ref{Temperature_Jump_RBC}), and (\ref{ICS}). Also, the heat source is considered as,
\begin{equation}
	\label{Heat_Source_P3}
	\
	f(x,t) =
	\begin{cases}
		10^{19} \hspace{2mm}(\hspace{1mm}\textrm{J}\textup{s}^{-1}\textup{m}^{-3}) , & 0\leqslant x\leqslant \frac{L_z}{5L_c},\hspace{3mm} 0\leqslant t\leqslant \bar{T}\\
		0. & \frac{L_z}{5L_c}< x\leqslant \frac{L_z}{L_c},\hspace{3mm} 0\leqslant t\leqslant \bar{T}\\
	\end{cases}
	\
\end{equation}
In order to numerically investigate the problem, parameters $B=0.05$, $Kn=10$ and $\lambda=\sqrt{0.0037+0.4022e^{-Kn}}$ are taken from Ghazanfarian and Abbassi \cite{ghazanfarian2009effect}. Also $\phi_1(t)$, $\phi_2(t)$, $\psi_1(x)$, and $\psi_2(x)$ are zero and the maximum numerical error and convergence order for $\bar{T}=2$ are presented in Tab. \ref{Tabel_P3_2}. As it is obvious, by halving the space and time step, maximum numerical error decreases and the convergence order tends to $O(h)$ as expected. As reported in Tab. \ref{Tabel_P3_2}, the average relative error of the maximum numerical error and the convergence order at $Kn=10$ is almost zero and the obtained numerical results are in good agreement with that of Ji \emph{et al} \cite{ji2018numerical2}. The temperature profile for different values of $\alpha$ at Kn=10 when t=200 ps are shown in Fig. \ref{case3}. As it is seen in Fig. \ref{case3}(a), when $t=200 ps$, for $\alpha$ being 0.3, 0.5, 0.7 and 0.9, the maximum temperature is respectively, 304.8, 305.4, 305.5 and 305.6 K. Also, Fig. \ref{case3}(b) demonstrates the temperature distribution at $X=7.5 nm$ for different values of $\alpha$ and Kn=10. Evidently, for $\alpha$ being 0.7 and 0.9 there exist oscillations at the beginning of the plot that indicates temperature instability.
	\begin{table}[htbp]
	\caption{The maximum value of the numerical error and the convergence order for $\textit{Kn}=10$}
	\label{Tabel_P3_2}
	\begin{center}
		\begin{small}
			\begin{tabular}{llcllcll}
				\hline
				\multicolumn{2}{c}{Ji et al.~\cite{ji2018numerical2}} &&\multicolumn{2}{c}{Present results} &&$\textit{h}$&$\alpha$\\ [0.5ex]
				\cline{1-2}
				\cline{4-5}
				\multicolumn{2}{l}{$H_{\infty}(h)\hspace{15mm}Rate_2$}&&\multicolumn{2}{l}{$H_{\infty}(h)\hspace{15mm}Rate_2$}&&&\\
				\hline
				&$1.460\times 10^{-3}$&&&$1.460\times 10^{-3}$&&$0.05$&$0.3$\\
				$1.099$&$6.816\times 10^{-4}$&&$1.099$&$6.816\times 10^{-4}$&&$0.025$&\\
				$1.063$&$3.263\times 10^{-4}$&&$1.063$&$3.263\times 10^{-4}$&&$0.0125$&\\
				$1.000$&$1.632\times 10^{-4}$&&$1.000$&$1.632\times 10^{-4}$&&$0.00625$&\\
				&$1.571\times 10^{-3}$&&&$1.571\times 10^{-3}$&&$0.05$&$0.5$\\
				$1.066$&$7.504\times 10^{-4}$&&$1.066$&$7.504\times 10^{-4}$&&$0.025$&\\
				$0.983$&$3.797\times 10^{-4}$&&$0.983$&$3.797\times 10^{-4}$&&$0.0125$&\\
				$1.002$&$1.896\times 10^{-4}$&&$1.002$&$1.896\times 10^{-4}$&&$0.00625$&\\
				&$1.644\times 10^{-3}$&&&$1.644\times 10^{-3}$&&$0.05$&$0.7$\\
				$1.021$&$8.100\times 10^{-4}$&&$1.021$&$8.100\times 10^{-4}$&&$0.025$&\\
				$1.004$&$4.040\times 10^{-4}$&&$1.004$&$4.040\times 10^{-4}$&&$0.0125$&\\
				$1.001$&$2.018\times 10^{-4}$&&$1.001$&$2.018\times 10^{-4}$&&$0.00625$&\\
				\hline
			\end{tabular}
		\end{small}
	\end{center}
\end{table}

\subsection{Verification: Case four}
\begin{figure}[htbp]
	\begin{center}
		\includegraphics[scale=0.5]{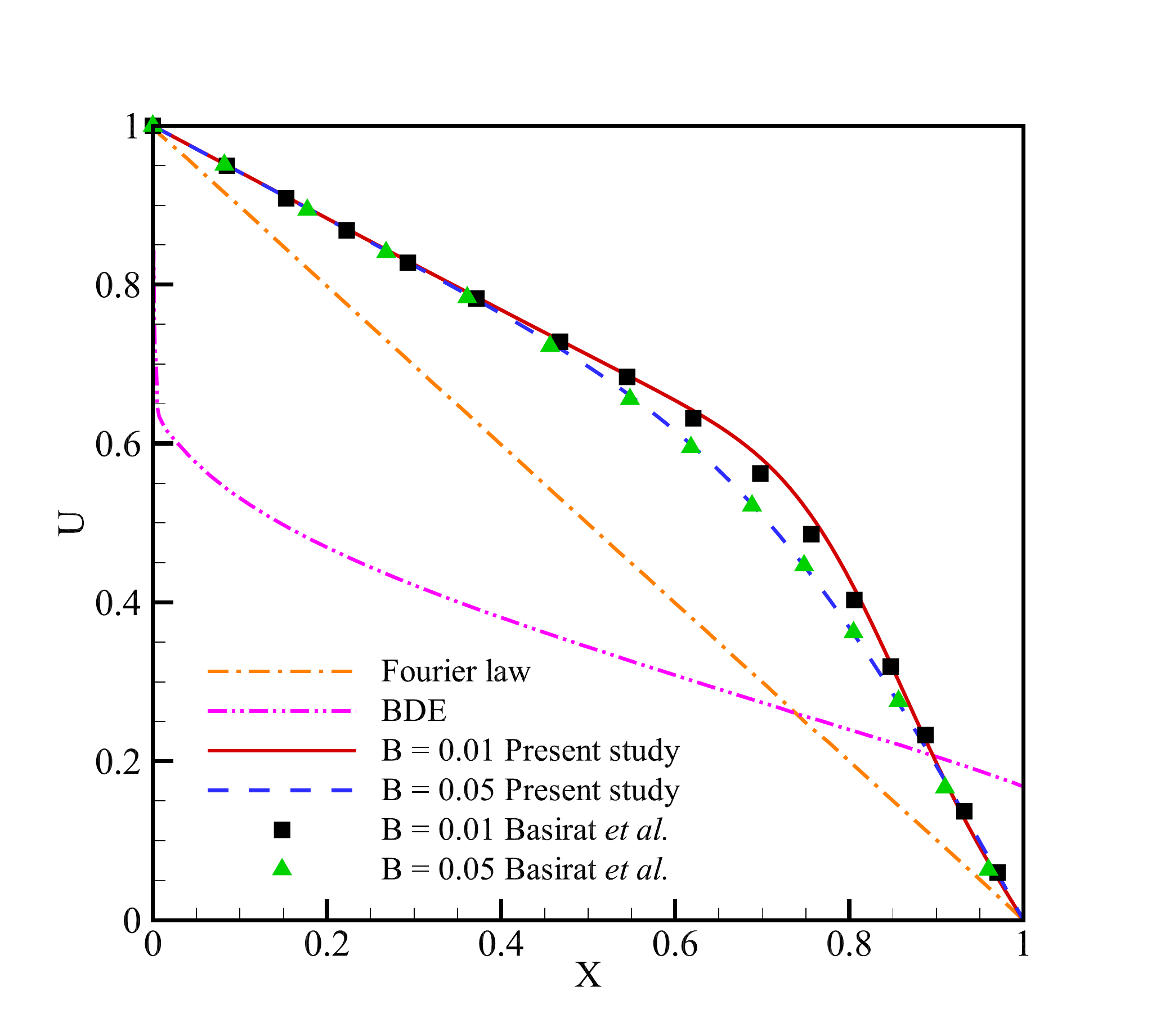}
	\end{center}
	\caption{Comparison of the normalized temperature distribution using the Fourier's law, the BDE model, the DPL model~\cite{basirat2006implementation}, and the fractional DPL model at $Kn=1$ and $\bar{T}=2$.}
	\label{Fig.3.6}
\end{figure}

For the fourth case of verification, the heat transfer at a one-dimensional slab without a heat source considering the Dirichlet boundary condition, as shown in Fig. \ref{CasesSchematic} (d), is studied. The case is investigated for different values of the Knudsen numbers and the obtained results are compared with the data available from Basirat and Ghazanfarian \cite{basirat2006implementation}. Here, the governing equation is Eq. \ref{DFDPL_eq} as well. Also the boundary and initial conditions are,
\begin{align}
	\nonumber
	&u(0,t)=\left ( T_w-T_0 \right )/T_0,\hspace{3mm} 0< t\leqslant \bar{T}\\
	\nonumber
	&u(1,t)=0,\hspace{3mm} 0< t\leqslant \bar{T}\\
	\label{Basirat_ICS}
	&u(x,0)=0,\hspace{3mm} u_t(x,0)=0,\hspace{3mm} 0\leqslant x\leqslant 1.
\end{align}
We have considered T$_0$=300 K and T$_w$=360 K. The non-dimensional temperature $U=\frac{T-T_0}{T_w-T_0}$ for Kn=1 is plotted in Fig. \ref{Fig.3.6}. The comparison between the obtained non-dimensional temperature profile and the data available from the Fourier, the BDE equations, and the DPL models \cite{basirat2006implementation}, when Kn=1 and $\bar{T}=2$, is carried out. We have taken $\alpha \approx1$ in our developed fractional DPL scheme. Fig. \ref{Fig.3.6} presents an appropriate consistency with our results and the one appeared in \cite{basirat2006implementation}. The results present a sudden drop in temperature in the temperature distribution profile. So the obtained results present better consistency with that of the BDE model in points near to the bottom boundary. The average relative error between our results and the one obtained by Basirat \emph{et al.} when Kn=1, $\bar{T}=2$, is less than 3$\%$ and 5$\%$, respectively, for $B$=0.01 and $B$=0.05.
\subsection{Verification: Case five}
\begin{figure}[htbp]
	\begin{center}
		\includegraphics[scale=0.4]{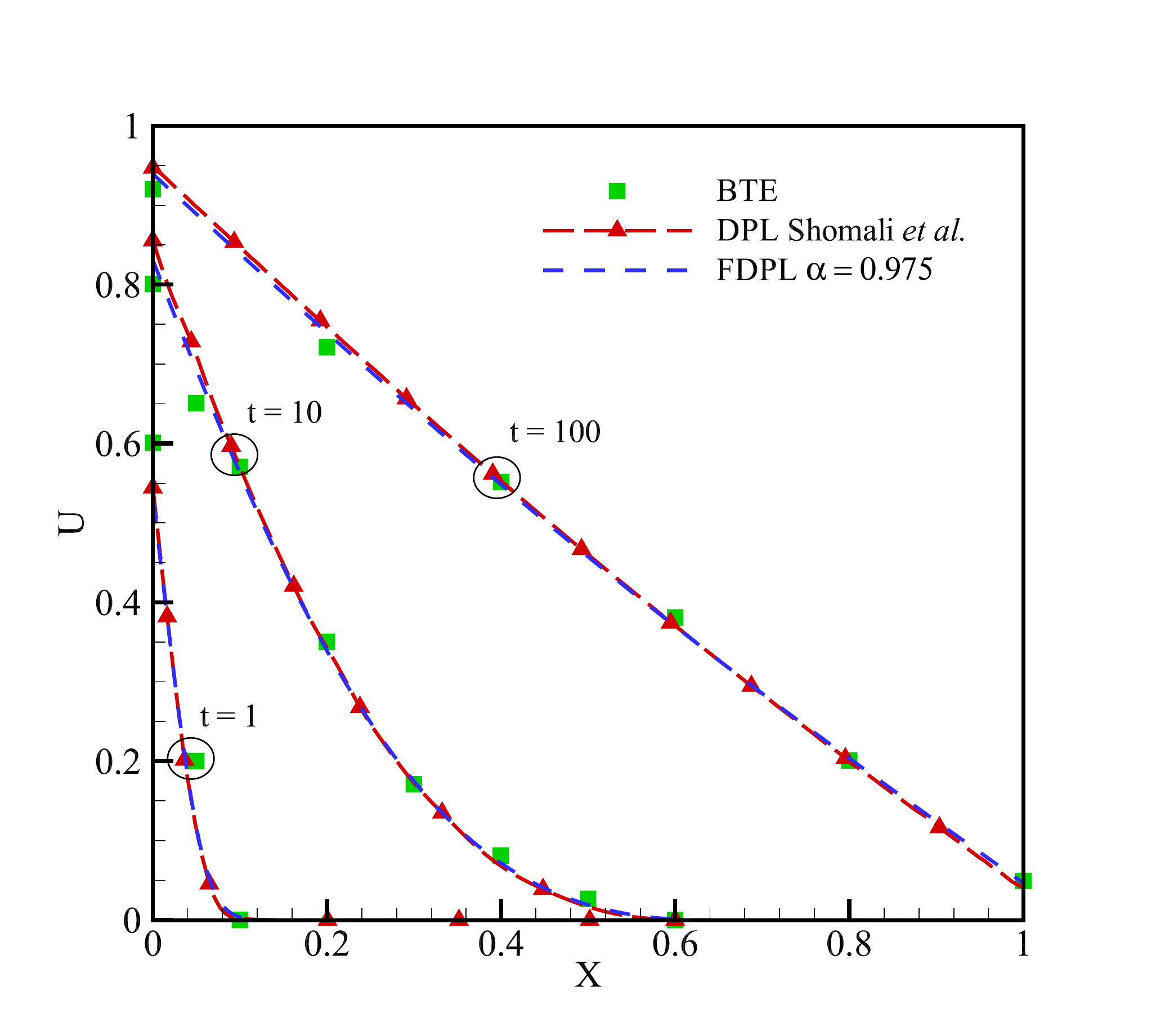}
	\end{center}
	\caption{The non-dimensional temperature distribution obtained from Boltzmann transport equation, the DPL model~\cite{shomali2015effect}, and fractional DPL at $Kn=0.1$}
	\label{Fig.3.7}
\end{figure}

As Fig.\ref{CasesSchematic} (e) suggests, the verification case number five deals with investigation of the one-dimensional silicon slab without heat source, including the temperature jump boundary condition. The results are verified with the data in \cite{shomali2015effect}. Here, the initial temperature of the silicon slab is T$_0$=300 K and its length is L$_z$=1000 nm. As there exists no heat source, the characteristic length is equal to the slab length. Also, the thermal properties are presented in Tab. \ref{Tabel_P3_1}. In order to investigate the temperature profile in the silicon slab, the top temperature is increased up to T$_w$=360 K while the bottom boundary is kept at the environment temperature. As previously mentioned, the value of parameters $B$ and $\lambda$ along with the order of the fractional DPL model $\alpha$, for each time, should be found such that the results obtained from the fractional DPL coincides with that of the Boltzmann equation. Determining several different constants for various cases is not desirable as using them is not easy. Here, the following parameters for $B$ and $\lambda$ are obtained,
\begin{align}
	\nonumber
	\
	B &=
	\begin{cases}
		0.1t\left [ \Gamma(1+\alpha) \right ]^{-1/\alpha}, & t\leqslant \left [ \Gamma(1+\alpha) \right ]^{1/\alpha}\\
		0 , & t>\left [ \Gamma(1+\alpha) \right ]^{1/\alpha}\\
	\end{cases}
	\ \\
	\label{lambda_eq}
	\
	\lambda &=
	\begin{cases}
		0.7t\left [ \Gamma(1+\alpha) \right ]^{-1/\alpha} , & t\leqslant 0.1\left [ \Gamma(1+\alpha) \right ]^{1/\alpha}\\
		0.5, & t> 0.1\left [ \Gamma(1+\alpha) \right ]^{1/\alpha}\\
	\end{cases}
	\
\end{align}
Also, the parameter $\alpha$=0.975 is found to show better consistency of the results. As it is obvious, reaching the steady state, $B$ and $\lambda$ become constant. Due to slow penetration of the heat for case Kn=0.1, the non-dimensional temperature distributions for $\bar{T}$=1, $\bar{T}$=10, and $\bar{T}$=100 are plotted in Fig. \ref{Fig.3.7}. The justifiable consistency between the results calculated from fractional DPL and the available data from Shomali \emph{et al.} \cite{shomali2015effect} were found. While the average relative error between what obtained from the DPL and the Boltzmann models is about 13\%, this error is less than 11\% for the results calculated from the fractional DPL model and the Boltzmann equation. This suggests that the fractional DPL model along with the temperature jump boundary condition can predict more precious temperature profiles.
\section{Results and discussions}
\label{results}
The five cases of studies are presented in Fig. \ref{CasesSchematic} (f)-(j). The considered heat source and the boundary conditions are similar to what exists in a real transistor. The initial temperature of the MOSFET is taken to be the room temperature, $T_0=300 \ K$. As Fig. \ref{CasesSchematic} (f) shows, the first investigated geometry consists of a very thin silicon layer, which has a heat generation zone near the top boundary. The most generated heat is conducted towards the bottom boundary, and finally is discharged to the surrounding environment. This approximation is valid unless there exists no heat flow through the top boundary. In other words, due to creation of the oxide layer which has almost infinite thermal resistance on top of the MOSFET, the top boundary is considered isolating. The temperature at the bottom boundary is the room temperature and the heat generation zone of Q=10$^{19}$ W/$m^3$ and lengthvL$_h$=10 nm is contemplated. The length of the transistor is L$_z$=50 nm. This structure is investigated for two cases of (a) constant thermal properties, and (b) film-dependent thermal characteristics. The next studied geometry, Fig. \ref{CasesSchematic} (g) is a transistor with a buried silicon oxide on top of the silicon layer. Other conditions in this case are similar to that of the case in Fig. \ref{CasesSchematic} (f). In an additional step, the silicon channel is replaced with the newest proposed quasi-one-dimensional channels \cite{galiybuilding}. Two materials of In$_{4}$Se$_{3}$ and TiS$_{3}$ are taken into account. Such transistors shown in Figs. \ref{CasesSchematic} (i) and (j) are called quasi-one-dimensional transistors. Also, in order to investigate the functionality and reliability of these MOSFETs in comparison with the one including the silicon channel, the geometry in Fig. \ref{CasesSchematic} (h) is also studied. As the Figs. \ref{CasesSchematic} (h)-(j) suggests the temperature of the bottom boundary according to the room temperature and the temperature jump boundary condition is assumed. Also, Q=10$^{19}$ W/m$^3$ is located at the middle of the channel. The parameters L$_h$=10 nm and L$_z$=50 nm are also utilized. In these five studied geometries, L$_h$ is considered as the characteristic length. Also, $\lambda$ and $B$ are defined in Eq. \ref{lambda_eq}. Moreover, as previously mentioned for a solid material, the thermal conductivity is k=$\rho$ C$_p$$|v|\frac{\Lambda}{3}$. On the other hand, it should be noted that if the transistor length L$_z$ is much larger than the phonon mean free-path $\Lambda$, the limited size effects appear. Under this circumstance, the mean free path of phonons is calculated as \cite{schelling2002comparison}:
\begin{align}
	\label{Lambda_effective}
	\frac{1}{\Lambda_{eff}}=\frac{1}{\Lambda}+\frac{4}{L_z}.
\end{align}
As $\Lambda$ can be replaced from $\Lambda$=$|v|\tau_q$, the $\Lambda_{eff}$ is obtained as:
\begin{align}
	\label{Lambda_eff}
	\frac{1}{\Lambda_{eff}}=\frac{\rho C_p \left | v \right |}{3k}+\frac{4}{L_z}.
\end{align}
Also, the thermal conductivity is calculated via:
\begin{align}
	\label{k_film}
	\frac{1}{k_{film}}=\frac{1}{k}+\frac{12}{\rho C_p \left | v \right |L_z}.
\end{align}
Eq. \ref{k_film} suggests that the silicon film thermal conductivity is always less than that of the bulk silicon. Considering the size effect in nano dimension, for certain, makes the obtained results more reliable. Also, the thermal properties of silicon, silicon dioxide, TiS$_3$, and In$_4$Se$_3$ are presented in Tab. \ref{S_SO2}.
\begin{table}[htbp]
	\caption{Thermal properties of Si~\cite{yang2005simulation}, SiO$_2$~\cite{goodson1992effect}, TiS$_3$~\cite{abdulsalam2015structural,zhang2017titanium}, and In$_4$Se$_3$~\cite{luu2020origin}.}
	\label{S_SO2}
	\begin{center}
		\begin{small}
			\begin{tabular}{c@{\hspace{1cm}}l@{\hspace{1cm}}l@{\hspace{1cm}}l@{\hspace{1cm}}l}
				\hline
				&$\left | v \right |(\hspace{0mm}\textrm{m}\textup{s}^{-1})$&$k(\hspace{0mm}W\textup{m}^{-1}\textup{K}^{-1})$&$\rho C_p(\hspace{1mm}\textrm{J}\textup{m}^{-3}\textup{K}^{-1})$&$\Lambda(\textup{nm})$\\
				\hline
				$\textup{Si}$&$3000$&$150$&$1.5\times 10^{6}$&$100$\\
				$\textup{SiO}_2$&$5900$&$1.4$&$1.75\times 10^{6}$&$0.4$\\
				$\textup{TiS}_3$&$3535$&$8.28$&$3.877\times 10^{6}$&$1.81$\\
				$\textup{In}_4\textup{Se}_3$&$2010$&$0.9$&$1.505\times 10^{6}$&$0.892$\\
				\hline
			\end{tabular}
		\end{small}
	\end{center}
\end{table}
\subsection{Result: Cases (I) and (II)}
\begin{figure}[htbp]
	\begin{center}
		\includegraphics[scale=0.7]{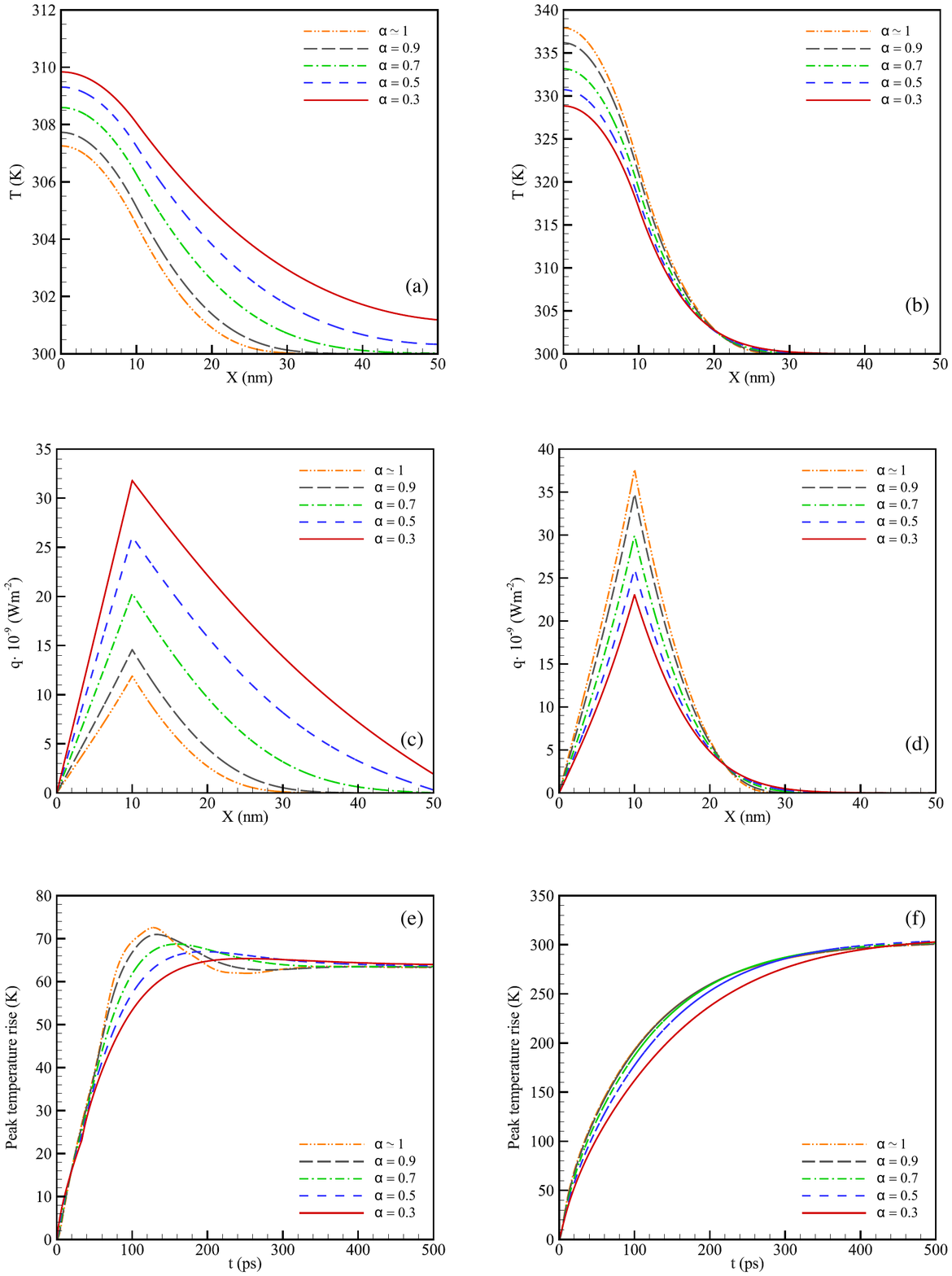}
	\end{center}
	\caption{The temperature profile, heat flux distribution and variation of the peak temperature for the case (I) considering the bulk thermal conductivity are subsequently presented in (a), (c), (e). Also, the figures (b), (d), (f), show the same as (a), (c) and (e) but for case (I) with the film thermal conductivity. The figures are plotted at t=10 ps.}
	\label{case6}
\end{figure}

In this section, the first case, Fig. \ref{CasesSchematic} (f), is studied while the bulk and also film thermal conductivity are taken into account. For all case studies, the parameters defined in Eq. \ref{lambda_eq} are utilized. Figs. \ref{case6} (a), (c), (e) show, respectively, the temperature and heat flux distributions at t=10, and the maximum temperature obtained from the fractional DPL model with considering the bulk thermal conductivity. Also, Figs. \ref{case6} (b), (d), (f) show the results when the film thermal characteristics are considered. As it is seen in Figs. \ref{case6} (a) and (b), considering size dependency of the thermal properties, changes the temperature distribution such that the temperature jump at the bottom boundary decreases. This decrement is due to the decrease of thermal conductivity from 150$Wm^{-1}K^{-1}$ to 16.67$Wm^{-1}K^{-1}$ which itself, results in reduction of the Knudsen number from 10 to 1.5. So, the temperature jump controlling term, $\Lambda Kn$ also decreases. As the thermal conductivity decreases, the total time in which all the transistors are affected by the heat generation zone increases. Consequently, the time for reaching the steady state condition augments. Moreover, Fig. \ref{case6} (a) suggests that as the $\alpha$ increases, the temperature jump at the bottom boundary decreases, and the whole temperature distribution plot is placed lower. This trend is recognizable for the heat flux plot. Further, for a silicon transistor with size dependent thermal characteristics where the Knudsen number decreases by 1.1, the temperature jump at the bottom boundary is not sensible for all values of $\alpha$. Increasing $\alpha$, the penetration heat decreases. In other words, when $\alpha$ reaches 1, the heat penetration becomes smaller. As the temperature jump reduces, for the similar time, the temperature in all positions takes higher values. The heat flux also obeys such behavior. It is worthy to mention that when the transistor size becomes much larger than the phonon mean free-path, the difference between the results appearing from taking film dependent or constant bulk thermal properties vanishes. It is obvious that the reason is reaching the bulk limit by increasing the system size. Additionally, the time dependency of the peak temperature rise is presented in Figs. \ref{case6} (e) and (f). It is found that the maximum temperature obtained from the fractional DPL model is always larger when the size dependent thermal properties are considered. For example, at t=40 ps and when $\alpha \simeq 1$, the maximum temperature rise is 32 K and 112 K, respectively for constant bulk and size-dependent thermal characteristics. On the other hand, the maximum temperature for the same system when considering both size and temperature dependent thermal properties, calculated from the common DPL model, is reported as 55 K and 112 K \cite{shomali2015effect}. Figure \ref{case6} (e) also shows that when $\alpha$ is equal to 0.7, 0.9, and almost 1, oscillations appear in maximum temperature plot which present the negative-bias temperature instability (NBTI) \cite{liao2014new,krishnan2003nbti,paul2005impact}. This phenomenon identified as the short-memory principle is observed in the works \cite{ji2019numerical,podlubny1998fractional}. In consequence, taking into account the bulk thermal properties for the film structures, notably underestimates the obtained temperature profiles and the peak temperature rise. So, it is important to consider the size-dependent thermal properties of the film structures, in order to calculate a more accurate temperature distribution.

\begin{figure}[htbp]
	\begin{center}
		\includegraphics[scale=0.7]{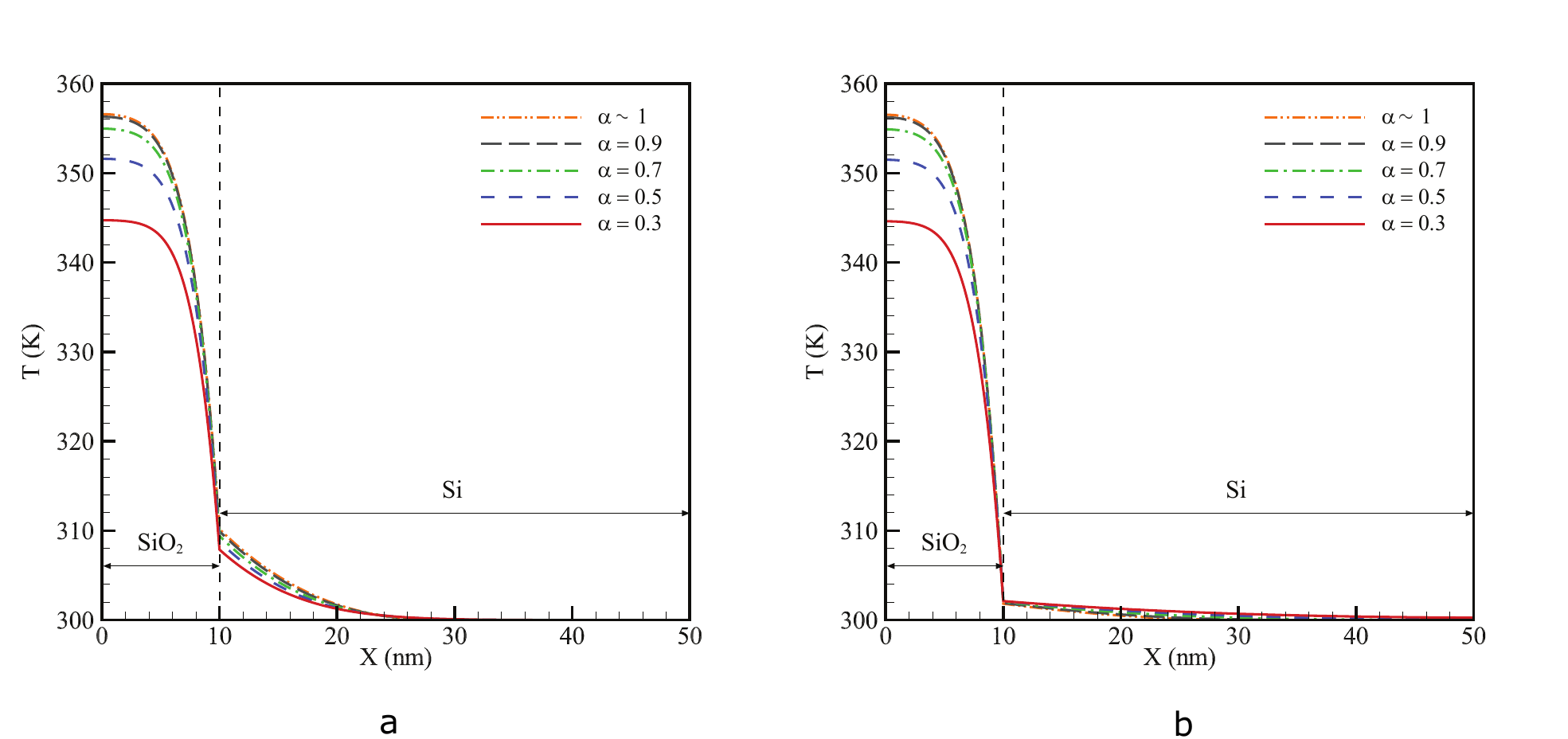}
	\end{center}
	\caption{Temperature distribution obtained from the fractional DPL model for the case (II) at 10 ps, (a) film properties, (b) bulk properties.}
	\label{case7}
\end{figure}

In the following, the results for the second case, Fig. \ref{CasesSchematic} (g), will be presented. Here, the heat generation zone is considered at the top of the transistor, at a place where there exists silicon dioxide. Using SiO$_2$ makes the effective electrical capacitance reduce, and it increases the operating speed \cite{sverdrup2001sub}. On the other hand, the temperature increase in the oxide layer which is due to the lower thermal conductivity coefficient of SiO$_2$ relative to the Si, can cause the transistor failure. In SiO$_2$/Si transistors, the silicon layer with high thermal conductivity increases the heat transfer, and helps the generated heat in the heating source to leave. So, the buried oxide layer operates like a cage for the heat such that the heat tends to flow from the silicon body. Figures. \ref{case7} (a) and (b) show the temperature profile for the SiO$_2$/Si transistors with bulk and size-dependent thermal properties. It is obvious that the low thermal conductivity of the SiO$_2$ has caused the temperature increase remarkably such that the temperature everywhere in the SiO$_2$/Si transistor is larger than that of Si transistor without any SiO$_2$ layer. At the same time, considering the limited size effect reduces the thermal conductivity of the SiO$_2$ from 1.4$Wm^{-1}K^{-1}$ to 1.35$Wm^{-1}K^{-1}$. This also causes the Knudsen number to decrease from 0.04 to 0.0394. So, considering the size-effects makes the thermal conductivity and the Knudsen number in the SiO$_2$ layer to change slightly. Consequently, As Figs. \ref{case7} (a) and (b) show, the low thermal conductivity of SiO$_2$ also makes the temperature distribution behavior very similar in SiO$_2$ parts of the transistor with both different defined thermal properties. Also, taking into account the size-effects in the silicon part of the transistor results in prominent decrement of thermal conductivity and Knudsen number, subsequently, from 150$Wm^{-1}K^{-1}$ to 16.67$Wm^{-1}K^{-1}$ and from 10 to 1.1. Accordingly, the temperature profile plots for the two cases with different bulk or size-dependent thermal properties are completely dissimilar. Besides, the size-dependent thermal specificities give rise to increment of the temperature at two-layer contact area, such that the temperature value for $\alpha \simeq 1$ increases from 301.8 K to 310.1 K.
\subsection{Result: Cases III-V}
\begin{figure}[htbp]
	\centering
	\includegraphics[scale=0.7]{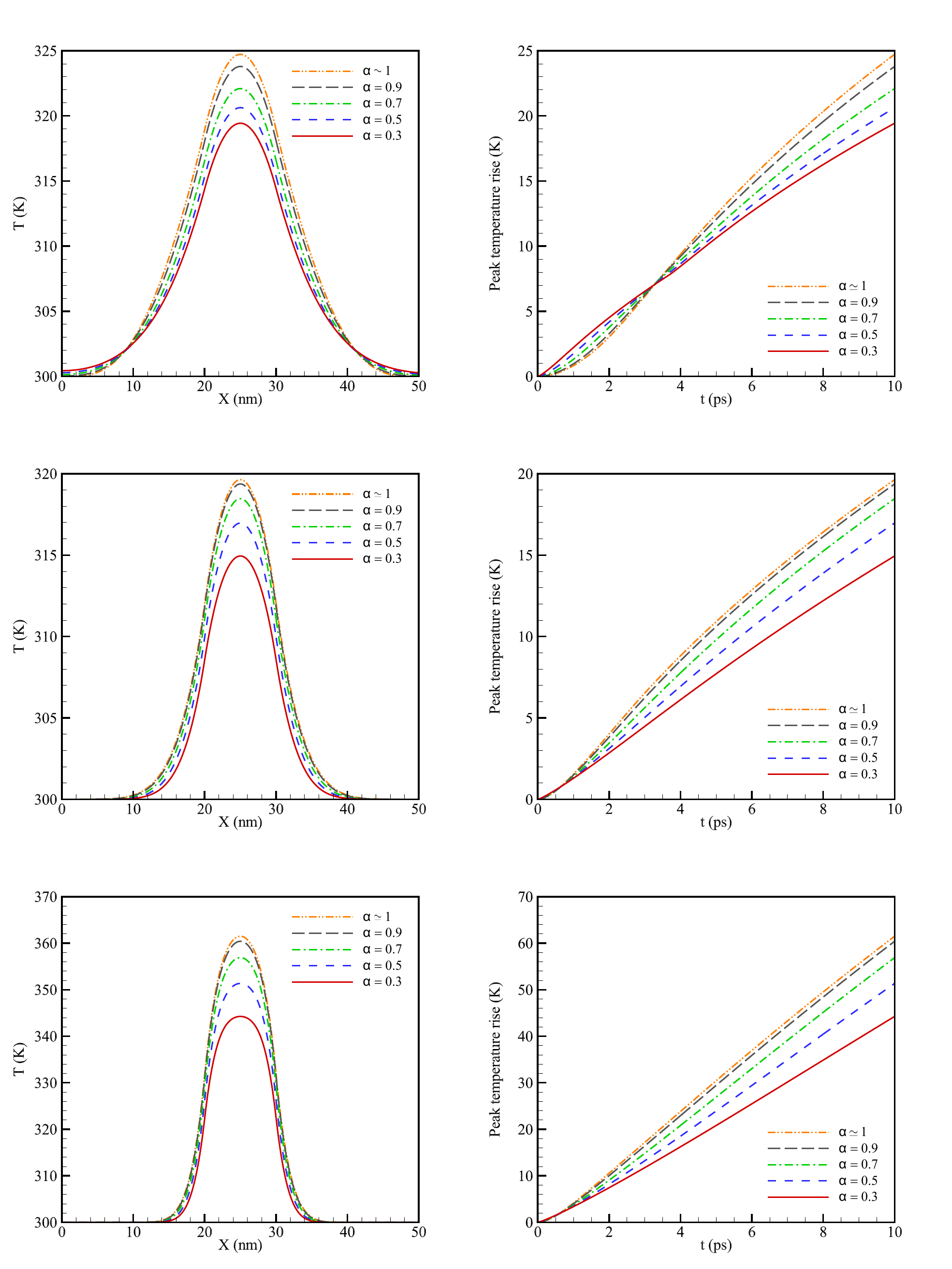}
	\caption{(a), (c) and (e), the temperature distribution at t=10 ps, respectively for, silicon, TiS$_3$, and In$_4$Se$_3$ quasi-one-dimensional transistors. (b), (d), and (f), the variation of the peak temperature in the, respectively, silicon, TiS$_3$, and In$_4$Se$_3$ transistors with film thermal properties.}
	\label{case8}
\end{figure}

\begin{figure}[htbp]
	\centering
	\includegraphics[scale=0.7]{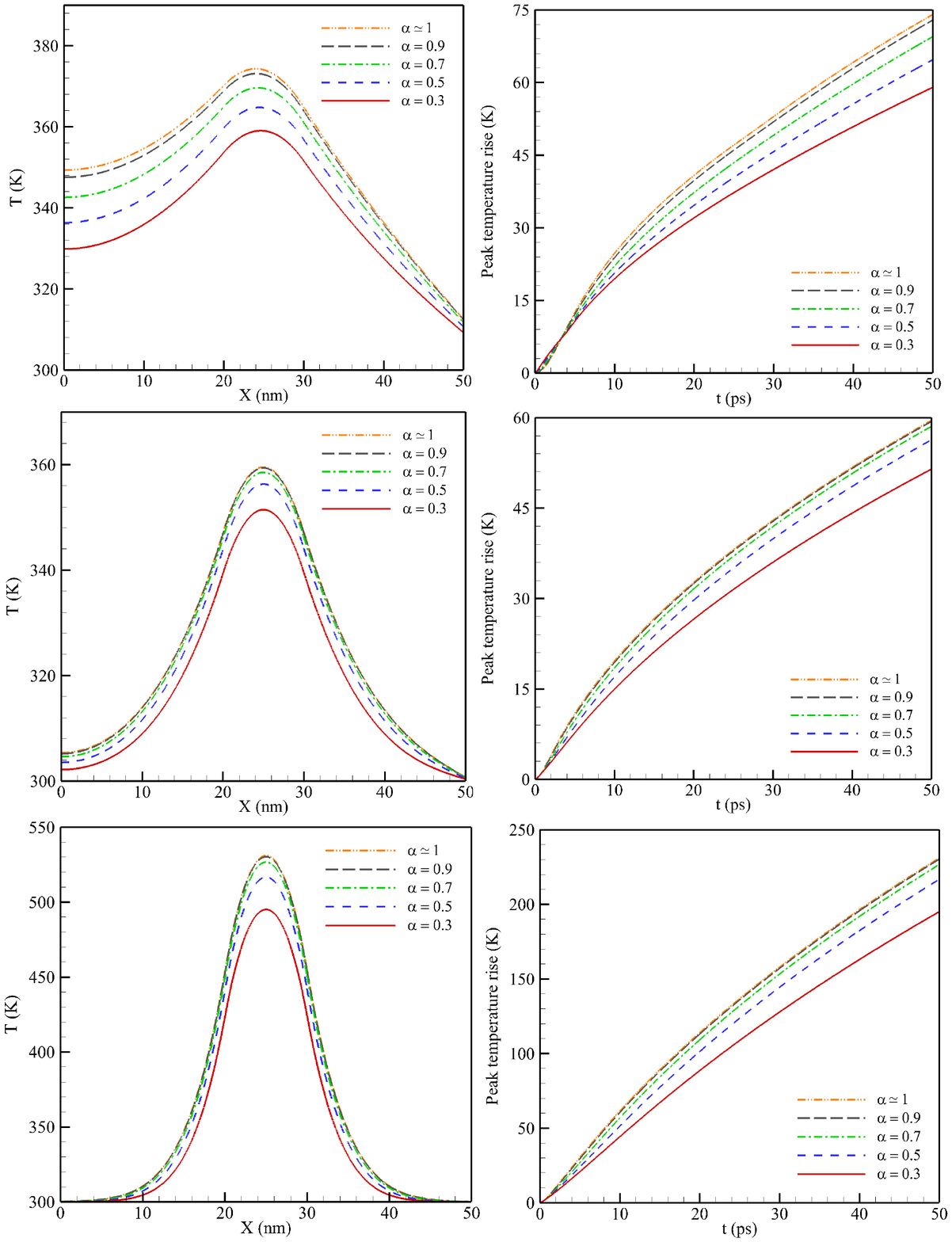}
	\caption{The same as Fig. \ref{case8}, but when t=50 ps.}
	\label{case8b}
\end{figure}

In this section, the obtained results for the cases presented in Figs. \ref{CasesSchematic} (h), (i), and (j), while considering the film-dependent thermal properties, are discussed. As previously mentioned, the heat generation zone is considered at the center of the transistor. In Figs. \ref{case8} (a), (c), and (e), the temperature distribution profiles obtained from fractional DPL model, respectively, for silicon, TiS$_3$, and In$_4$Se$_3$ at t=10 ps are given. It is obvious that the temperature of the silicon transistor presents a higher value all over the transistor relative to the TiS$_3$ and In$_4$Se$_3$ temperature, for all values of $\alpha$. Also, as $\alpha$ augments, the temperature jump value at the right boundary decreases while the temperature increases. Also, in similarity to the previously studied geometries, with $\alpha$ reduction, the heat penetration over the transistor length enhances. For $\alpha \simeq 1$, the maximum temperature rise for the silicon transistor at t=10, is 24.72 K. Further, as the Knudsen number is 1.1, reducing $\alpha$ and increasing the temperature jump at the bottom boundary, the peak temperature in early times (t$<$3.28 ps), has larger values. The temperature distribution profiles of TiS$_3$ and In$_4$Se$_3$, obtained from the fractional DPL model at t=10 ps for different values of $\alpha$, are respectively shown in Figs. \ref{case8} (c) and (e). It is obtained that for the same values of $\alpha$, the temperature profile for In$_4$Se$_3$ MOSFET is placed in a higher temperature range relative to TiS$_3$. This can be justified as the Knudsen number, the temperature jump, and the heat flux phase lag for In$_4$Se$_3$ (Kn=0.083 and $\tau_q$=0.414 ps) are smaller than that of TiS$_3$ ($\tau_q$=0.447 ps and Kn=0.158). Also as Fig. \ref{case8} (a) suggests, due to the larger value of the Knudsen number and the temperature jump for the silicon FET relative to the geometries presented in Fig. \ref{CasesSchematic} (i) and (j), the temperature distribution around boundaries, increases in value with $\alpha$ reduction. Also comparing Figs. \ref{case8} (a), (c), and (e), one finds the heat penetration in the silicon MOSFET is higher than in titanium trisulfide and tetraindium triselenide MOSFETs. Further, the localized heating is also obvious. The maximum temperature time-variation plot of TiS$_3$ and In$_4$Se$_3$ are demonstrated in Figs. \ref{case8} (d) and (f). It is shown that the peak temperature rise when $\alpha \simeq 1$, is 19.63 and 61.48, respectively, for TiS$_3$ and In$_4$Se$_3$. Conclusively, as the results for the maximum temperature of Si, TiS$_3$ and In$_4$Se$_3$ confirm, the TiS$_3$ FET having the least peak temperature rise, has the highest reliability. Also, as Fig. \ref{case8b} demonstrates, this trend is also seen for longer periods of time. So one can deduce that, the formed hot spot in the titanium trisulfide FET is cooler than the other two studied MOSFETs. So, the TiS$_3$ FET is suggested as the suitable replacement for the old-fashioned silicon transistors.

\section{Conclusions}
\label{con}
Due to the need for increasing the performance of electronic devices, reducing the sizes of these systems is inevitable. On the other hand, as the system characteristic length decreases, the Fourier law loses its validity. Instead, atomistic methods such as the molecular dynamics (MD) simulation or the phonon Boltzmann equation (PBE) are utilized. These methods have high computational costs, and also owing to their complexity, there exist limitations for applying them in complex structures. So, researchers with the help of precious results obtained from the atomistic methods can provide new models for modified classical equations like the DPL model. The new proposed models can produce accurate atomistic results with lower computational costs. In this regard, there have been many attempts to employ the DPL model for nanoscale problems. Recently, the combination of the fractional calculus and the phase lag theory, leading to more accurate results, has been the center of interest. On the other hand, the newly proposed TiS$_3$, and In$_4$Se$_3$ FETs are obtained to be suitable candidates for the silicon channel replacements. 

In the present study, the fractional DPL model discretized using the finite difference scheme is utilized to investigate the one-dimensional MOSFETs and quasi-one-dimensional FETs. To fulfill this aim, first, verification of the developed fractional DPL model has been performed considering five different cases. Then the elaborated fractional DPL method is applied for the newly quasi-one-dimensional suggested transistors while the thermal properties are taken to be size-dependent. It is obtained that considering size-dependent thermal characteristics make the peak temperature rise to increase, up to 250\%. In other words, considering the bulk thermal properties for film structures, remarkably underestimates the obtained temperature profiles and the temperature rise. So, it is important to consider the size-dependent thermal properties of the film structures, in order to obtain a more accurate temperature distribution. Also, it has reduced the temperature jump at the bottom boundary. The maximum temperature of the hotspot. The reliability has been investigated for the FET containing new two-dimensional proposed materials with quasi-one-dimensional band gap, named as TMT. It has been obtained that among the studied FETs, the transistor with titanium trisulfide channel presents the least temperature increase (19.63 K). Hence, TiS$_3$ can be suggested as the noteworthy substitution of the silicon channel. Also, considering $\alpha$=0.975 for Kn=0.1 is found to present better results.


\begin{thebibliography}{10}
	\expandafter\ifx\csname url\endcsname\relax
	\def\url#1{\texttt{#1}}\fi
	\expandafter\ifx\csname urlprefix\endcsname\relax\def\urlprefix{URL }\fi
	\expandafter\ifx\csname href\endcsname\relax
	\def\href#1#2{#2} \def\path#1{#1}\fi
	
	\bibitem{galiybuilding}
	P.~V. Galiy, M.~Randle, A.~Lipatov, L.~Wang, S.~Gilbert, N.~Vorobeva, A.~Kumar,
	C.-P. Kwan, J.~Nathawat, B.~Barut, et~al., Building the quasi one dimensional
	transistor from 2d materials, 2019 IEEE 2nd Ukraine Conference on Electrical
	and Computer Engineering (UKRCON) (2019) 679--682.
	
	\bibitem{Superlattice2018}
	. Kumar, M. M. Tripathi, R. Chaujar, Comprehensive analysis of sub-20 nm black phosphorus based junctionless-recessed channel mosfet for analog/rf applications, Superlattices and Microstructures 116 (2018) 171–180.
	
	\bibitem{Shomali2019A}
	Z.~Shomali, R.~Asgari, Effects of low-dimensional material channels on energy
	consumption of nano-devices, International Communications in Heat and Mass
	Transfer 94 (2018) 77--84.
	
	\bibitem{wimmer2008spin}
	M.~Wimmer, I.~Adagideli, S.~Berber, D.~Tom{\'a}nek, K.~Richter, Spin currents
	in rough graphene nanoribbons: Universal fluctuations and spin injection,
	Physical review letters 100~(17) (2008) 177207.
	
	\bibitem{TMTC}
	A.~Patra, C.~S. Rout, Anisotropic quasi-one-dimensional layered
	transition-metal trichalcogenides: synthesis, properties and applications,
	RSC Advances 10 (2020) 36413--36438.
	
	\bibitem{lipatov2018quasi}
	A.~Lipatov, M.~J. Loes, H.~Lu, J.~Dai, P.~Patoka, N.~S. Vorobeva, D.~S.
	Muratov, G.~Ulrich, B.~KÃ¤stner, A.~Hoehl, et~al., Quasi-1d tis3 nanoribbons:
	mechanical exfoliation and thickness-dependent raman spectroscopy, ACS nano
	12~(12) (2018) 12713--12720.
	
	\bibitem{tzou1995unified}
	D.~Y. Tzou, A unified field approach for heat conduction from macro-to
	micro-scales, J. Heat Transfer 117~(1) (1995) 8--16.
	
	\bibitem{tzou2001temperature}
	D.~Tzou, K.~Chiu, Temperature-dependent thermal lagging in ultrafast laser
	heating, International Journal of Heat and Mass Transfer 44~(9) (2001)
	1725--1734.

   \bibitem{Nasri20152}
	F. Nasri, M.F.B. Aissa, and H. Belmabrouk, Effect of second-order temperature jump in metal-oxide-semiconductor field effect transistor with dual-phase-lag model, Microelectronics Journal 46(1) (2015) 67--74.

    \bibitem{Nasri2015}
	F. Nasri, M.F.B. Aissa, M.H. Gazzah, H. Belmabrouk, 3D thermal conduction in a nanoscale Tri-Gate MOSFET based on single-phase-lag model, Applied Thermal Engineering, 91 (2015) 647--653.
	
	\bibitem{ghazanfarianshomali2012investigation}
	J.~Ghazanfarian, Z.~Shomali, Investigation of dual-phase-lag heat conduction
	model in a nanoscale metal-oxide-semiconductor field-effect transistor,
	International Journal of Heat and Mass Transfer 55~(21-22) (2012) 6231--6237.
	
	\bibitem{ghazanfarianabbassi2012investigation}
	J.~Ghazanfarian, A.~Abbassi, Investigation of 2d transient heat transfer under
	the effect of dual-phase-lag model in a nanoscale geometry, International
	Journal of Thermophysics 33~(3) (2012) 552--566.
	
	\bibitem{samian2013thermal}
	R.~Samian, A.~Abbassi, J.~Ghazanfarian, Thermal investigation of common 2d fets
	and new generation of 3d fets using boltzmann transport equation in
	nanoscale, International Journal of Modern Physics C 24~(9) (2013) 1350064.
	
	\bibitem{samian2014transient}
	R.~Samian, A.~Abbassi, J.~Ghazanfarian, Transient conduction simulation of a
	nano-scale hotspot using finite volume lattice boltzmann method,
	International Journal of Modern Physics C 25~(4) (2014) 1350103.
	
	\bibitem{Moghaddam2014}
	M.~Moghaddam, J.~Ghazanfarian, A.~Abbassi, Implementation of dpl-dd model for
	the simulation of nanoscale mos devices, IEEE Transactions on Electron
	Devices 61 (2014) 3131--3138.
	
	\bibitem{shomali2014investigation}
	Z.~Shomali, A.~Abbassi, Investigation of highly non-linear dual-phase-lag model
	in nanoscale solid argon with temperature-dependent properties, International
	Journal of Thermal Sciences 83 (2014) 56--67.
	
	\bibitem{shomali2015effect}
	Z.~Shomali, J.~Ghazanfarian, A.~Abbassi, Effect of film properties for
	non-linear dpl model in a nanoscale mosfet with high-k material:
	Zro$_2$/hfo$_2$/la$_2$o$_3$, Superlattices and Microstructures 83 (2015)
	699--718.
		
	\bibitem{Rev2015}
	J.~Ghazanfarian, Z.~Shomali, A.~Abbassi, Macro- to nanoscale heat and mass
	transfer: The lagging behavior, Internatinal Journal of Thermophysics 36
	(2015) 1416--1467.
	
	\bibitem{saghatchi2015}
	r.~Saghatchi, J.~Ghazanfarian, A novel sph method for the solution of
	dual-phase-lag model with temperature-jump boundary condition in nanoscale,
	Applied Mathematical Modelling 39 (2015) 1063--1073.
	
	\bibitem{Shomali2016}
	Z.~Shomali, A.~Abbassi, J.~Ghazanfarian, Development of non-fourier thermal
	attitude for three-dimensional and graphene-based mos devices, Applied
	Thermal Engineering 104 (2016) 616--627.
	
    \bibitem{Shiomi2016}
    J.~Shiomi and S.~Maruyama, Non-Fourier heat conduction in a single-walled carbon nanotube: classical molecular dynamics simulations, Phys. Rev. B, 73 (2006) 205420.
		
\bibitem{Nasri2017}
		F. Nasri, M.F.B. Aissa, and H. Belmabrouk, Nanoheat conduction performance of black phosphorus field-effect transistor, IEEE Transactions on Electron Devices 64(6) (2017) 2765--2769.
	
	\bibitem{Pedar2017}
	Z.~Shomali, B.~Pedar, J.~Ghazanfarian, A.~Abbassi, Monte-carlo parallel
	simulation of phonon transport for 3d silicon nano-devices, International
	Journal of Thermal Sciences 114 (2017) 139--154.
	
	\bibitem{shomali2017cht}
	Z.~Shomali, J.~Ghazanfarian, A.~Abbassi, 3-d atomistic investigation of silicon
	mosfets, in: In Proceedings of CHT-17 ICHMT International Symposium on
	Advances in Computational Heat Transfer. Begel House Inc, 2017, pp.
	1385--1401.
	
	\bibitem{ghazanfarian2018}
	M.~Jamshidi, J.~Ghazanfarian, Dual-phase-lag analysis of cnt-mos2-zro2-sio2-si
	nano-transistor and arteriole in multi-layered skin, Applied Mathematical
	Modelling 60 (2018) 490--507.
	
	\bibitem{Zobiri2021}
	O. Zobiri, A. Atia, M. Arıcı, A three-dimensional analysis of heat trans- fer based on mesoscopic method in nanoscale si-mosfet and gr-fet, Su- perlattices and Microstructures (2021) 107123.
	
	\bibitem{Caputo1971}
	C.~M., M.~F, A new dissipation model based on memory mechanism, Pure and
	Applied Geophysics 91 (1971) 134--147.
	
	\bibitem{Caputo1971-2}
	C.~M., M.~F, Linear model of dissipation in anelastic solids, Rivis ta del
	Nuovo cimento 1 (1971) 161--198.
	
	\bibitem{Caputo1974}
	C.~M., Vibrations on an infinite viscoelastic layer with a dissipative memory,
	Journal of the Acoustical Society of America 56 (1974) 897--904.
	
	\bibitem{sherief2010fractional}
	H.~H. Sherief, A.~El-Sayed, A.~Abd El-Latief, Fractional order theory of
	thermoelasticity, International Journal of Solids and structures 47~(2)
	(2010) 269--275.
	
	\bibitem{mishra2016numerical}
	T.~Mishra, K.~Rai, Numerical solution of fspl heat conduction equation for
	analysis of thermal propagation, Applied Mathematics and Computation 273
	(2016) 1006--1017.
	
	\bibitem{ji2018numerical2}
	C.-C. Ji, W.~Dai, Z.-Z. Sun, Numerical method for solving the time-fractional
	dual-phase-lagging heat conduction equation with the temperature-jump
	boundary condition, Journal of Scientific Computing 75~(3) (2018) 1307--1336.
	
	\bibitem{ji2019numerical}
	C.-C. Ji, W.~Dai, Z.-Z. Sun, Numerical schemes for solving the time-fractional
	dual-phase-lagging heat conduction model in a double-layered nanoscale thin
	film, Journal of Scientific Computing 81~(3) (2019) 1767--1800.
	
	\bibitem{ghazanfarian2009effect}
	J.~Ghazanfarian, A.~Abbassi, Effect of boundary phonon scattering on
	dual-phase-lag model to simulate micro-and nano-scale heat conduction,
	International Journal of Heat and Mass Transfer 52~(15-16) (2009) 3706--3711.
	
	\bibitem{dai2013accurate}
	W.~Dai, F.~Han, Z.~Sun, Accurate numerical method for solving
	dual-phase-lagging equation with temperature jump boundary condition in nano
	heat conduction, International Journal of Heat and Mass Transfer 64 (2013)
	966--975.
	
	\bibitem{dai2002approximate}
	W.~Dai, R.~Nassar, An approximate analytic method for solving 1d
	dual-phase-lagging heat transport equations, International journal of heat
	and mass transfer 45~(8) (2002) 1585--1593.
	
	\bibitem{caputo1967linear}
	M.~Caputo, Linear models of dissipation whose q is almost frequency
	independent-II, Geophysical Journal International 13~(5) (1967) 529--539.
	
	\bibitem{basirat2006implementation}
	H.~Basirat, J.~Ghazanfarian, P.~Forooghi, Implementation of dual-phase-lag
	model at different knudsen numbers within slab heat transfer, Proc. of Int.
	Conf. on Modeling and Simulation (MS06) (2006) 895--899.
	
	\bibitem{schelling2002comparison}
	P.~K. Schelling, S.~R. Phillpot, P.~Keblinski, Comparison of atomic-level
	simulation methods for computing thermal conductivity, Physical Review B
	65~(14) (2002) 144306.
	
	\bibitem{yang2005simulation}
	R.~Yang, G.~Chen, M.~Laroche, Y.~Taur, Simulation of nanoscale multidimensional
	transient heat conduction problems using ballistic-diffusive equations and
	phonon boltzmann equation, J. Heat Transfer 127~(3) (2005) 298--306.
	
	\bibitem{goodson1992effect}
	K.~Goodson, M.~Flik, Effect of microscale thermal conduction on the packing
	limit of silicon-on-insulator electronic devices, IEEE Transactions on
	Components Hybrids and Manufacturing Technology 15~(5) (1992) 715--722.
	
	\bibitem{abdulsalam2015structural}
	M.~Abdulsalam, D.~P. Joubert, Structural and electronic properties of mx 3 (m=
	ti, zr and hf; x= s, se, te) from first principles calculations, The European
	Physical Journal B 88~(7) (2015) 1--11.
	
	\bibitem{zhang2017titanium}
	J.~Zhang, X.~Liu, Y.~Wen, L.~Shi, R.~Chen, H.~Liu, B.~Shan, Titanium trisulfide
	monolayer as a potential thermoelectric material: a first-principles-based
	boltzmann transport study, ACS applied materials \& interfaces 9~(3) (2017)
	2509--2515.
	
	\bibitem{luu2020origin}
	S.~D. Luu, A.~R. Supka, V.~H. Nguyen, D.-V.~N. Vo, N.~T.~Hung, K.~T.
	Wojciechowski, M.~Fornari, P.~Vaqueiro, Origin of low thermal conductivity in
	in4se3, ACS Applied Energy Materials 3~(12) (2020) 12549--12556.
	
	\bibitem{liao2014new}
	M.~Liao, Z.~Gan, New insight on negative bias temperature instability
	degradation with drain bias of 28 nm high-k metal gate p-mosfet devices,
	Microelectronics Reliability 54~(11) (2014) 2378--2382.
	
	\bibitem{krishnan2003nbti}
	A.~T. Krishnan, V.~Reddy, S.~Chakravarthi, J.~Rodriguez, S.~John, S.~Krishnan,
	Nbti impact on transistor and circuit: models, mechanisms and scaling effects
	[mosfets], in: IEEE international electron devices meeting 2003, 2003, pp.
	14--5.
	
	\bibitem{paul2005impact}
	B.~C. Paul, K.~Kang, H.~Kufluoglu, M.~A. Alam, K.~Roy, Impact of nbti on the
	temporal performance degradation of digital circuits, IEEE Electron Device
	Letters 26~(8) (2005) 560--562.
	
	\bibitem{podlubny1998fractional}
	I.~Podlubny, Fractional differential equations: an introduction to fractional
	derivatives, fractional differential equations, to methods of their solution
	and some of their applications, Elsevier, 1998.
	
	\bibitem{sverdrup2001sub}
	P.~G. Sverdrup, Y.~Sungtaek~Ju, K.~E. Goodson, Sub-continuum simulations of
	heat conduction in silicon-on-insulator transistors, J. Heat Transfer 123~(1)
	(2001) 130--137.
	
\end{thebibliography}
\end{document}